\documentclass[11pt]{article}
\usepackage{epsfig,color,amsmath,amssymb,euscript}
 \usepackage{mathabx}
\usepackage{fancyhdr}
\usepackage{rotating}
\usepackage{amsmath}
\usepackage{amsfonts}

\usepackage{setspace}
\usepackage{fancyhdr}
\usepackage[top=0.9in,bottom=0.9in,left=0.8in,right=0.8in]{geometry}
\usepackage{lscape}
\usepackage{amsthm}
\usepackage{enumerate}
\usepackage{threeparttable}
\usepackage{diagbox}
\usepackage{multirow}
\usepackage{makecell}
\usepackage{xcolor}
\usepackage{graphicx}
\usepackage[authoryear,round]{natbib}
\bibpunct{(}{)}{;}{a}{}{,}
\usepackage{enumerate}
\usepackage{mathrsfs}
\RequirePackage{hyperref}
\usepackage{doi}

 \usepackage{graphicx}
 \usepackage{epstopdf}

\RequirePackage{fancyvrb}

\usepackage{pifont}

\usepackage{setspace}
\usepackage{threeparttable}
\usepackage{caption}
\usepackage{subfigure}
\usepackage{authblk}
\usepackage{url}
\usepackage{framed}

\usepackage{algorithm}
\usepackage{algorithmic}
\usepackage{float}  
\usepackage{lipsum}


\makeatletter
\newcommand{\fdsy@scale}{1.0}
\newcommand\fdsy@mweight@normal{Book}
\newcommand\fdsy@mweight@small{Book}
\newcommand\fdsy@bweight@normal{Medium}
\newcommand\fdsy@bweight@small{Medium}
\newcommand{\pkg}[1]{{\fontseries{m}\fontseries{b}\selectfont #1}}
\let\proglang=\textsf
\newcommand\code{\bgroup\@makeother\_\@makeother\~\@makeother\$\@codex}
\def\@codex#1{{\normalfont\ttfamily\hyphenchar\font=-1 #1}\egroup}

\DefineVerbatimEnvironment{Code}{Verbatim}{}
\DefineVerbatimEnvironment{CodeInput}{Verbatim}{fontshape=sl}
\DefineVerbatimEnvironment{CodeOutput}{Verbatim}{}
\newenvironment{CodeChunk}{}{}

\DeclareFontFamily{U}{FdSymbolC}{}

\DeclareFontShape{U}{FdSymbolC}{m}{n}{
	<-7.1> s * [\fdsy@scale] FdSymbolC-\fdsy@mweight@small
	<7.1-> s * [\fdsy@scale] FdSymbolC-\fdsy@mweight@normal
}{}
\DeclareFontShape{U}{FdSymbolC}{b}{n}{
	<-7.1> s * [\fdsy@scale] FdSymbolC-\fdsy@bweight@small
	<7.1-> s * [\fdsy@scale] FdSymbolC-\fdsy@bweight@normal
}{}

\DeclareSymbolFont{arrows}{U}{FdSymbolC}{m}{n}
\SetSymbolFont{arrows}{bold}{U}{FdSymbolC}{b}{n}

\DeclareMathSymbol{\upvDash}{\mathrel}{arrows}{233}

\DeclareMathSymbol{\upmodels}{\mathrel}{arrows}{237}
\makeatother

\theoremstyle{definition}

\newtheorem{extension}{Extension}[section]

\newtheorem{method}{Method}[subsection]

\def\T{{\mathrm{\scriptscriptstyle \top} }}

\newcommand{\bA}{{\mathbf A}}
\newcommand{\bB}{{\mathbf B}}

\newcommand{\bD}{{\mathbf D}}

\newcommand{\bH}{{\mathbf H}}
\newcommand{\bI}{{\mathbf I}}
\newcommand{\bJ}{{\mathbf J}}
\newcommand{\bK}{{\mathbf K}}

\newcommand{\bM}{{\mathbf M}}

\newcommand{\bP}{{\mathbf P}}
\newcommand{\bQ}{{\mathbf Q}}

\newcommand{\bU}{{\mathbf U}}
\newcommand{\bV}{{\mathbf V}}
\newcommand{\bW}{{\mathbf W}}
\newcommand{\bX}{{\mathbf X}}
\newcommand{\bY}{{\mathbf Y}}
\newcommand{\bZ}{{\mathbf Z}}
\newcommand{\ba}{{\mathbf a}}
\newcommand{\bb}{{\mathbf b}}

\newcommand{\bu}{{\mathbf u}}
\newcommand{\bv}{{\mathbf v}}
\newcommand{\bw}{{\mathbf w}}
\newcommand{\bx}{{\mathbf x}}
\newcommand{\by}{{\mathbf y}}
\newcommand{\bz}{{\mathbf z}}
\newcommand{\bzero}{{\mathbf 0}}

\newcommand{\bbeta}  {\boldsymbol{\beta}}
\newcommand{\bfeta}  {\boldsymbol{\eta}}

\newcommand{\bepsilonb}{\boldsymbol{\varepsilon}}

\newcommand{\bOmega}{\boldsymbol{\Omega}}
\newcommand{\bTheta}{\boldsymbol{\Theta}}

\newcommand{\bSigma}{\boldsymbol{\Sigma}}

\newcommand{\bgamma}{\boldsymbol{\Gamma}}

\def\T{\scriptscriptstyle\top}

\theoremstyle{plain}

\makeatletter

\newcommand{\Date}[1]{\def\@Date{#1}}
\def\today{\number\day~\ifcase\month\or
 January\or February\or March\or April\or May\or June\or
 July\or August\or September\or October\or November\or December\fi~\number\year}

\allowdisplaybreaks[3]

\begin{document}


\title{HDTSA: An \proglang{R} Package for High-Dimensional Time Series Analysis}



\author{Jinyuan Chang $^{\rm a,b}$ \qquad Jing He $^{\rm a}$ \qquad Chen Lin $^{\rm a}$ \qquad Qiwei Yao $^{\rm c}$\\
$^{\rm a}$ Joint Laboratory of Data Science and Business
Intelligence, Southwestern University of Finance and Economics, Chengdu, China\\
$^{\rm b}$ Academy of Mathematics and Systems Science, Chinese Academy of Sciences, Beijing, China\\
$^{\rm c}$ Department of Statistics, The London School of Economics and Political Science, London, U.K.
}

\maketitle

\begin{abstract}
High-dimensional time series analysis has become increasingly important in fields such as finance, economics, and biology. The two primary tasks for high-dimensional time series analysis are modeling and statistical inference, which aim to capture the underlying dynamic structure and investigate valuable information in the data. This paper presents the \pkg{HDTSA} package for \proglang{R}, which provides a general framework for analyzing high-dimensional time series data. This package includes four dimension reduction methods for modeling: factor models, principal component analysis, CP-decomposition, and cointegration analysis. It also implements two recently proposed white noise test and martingale difference test in high-dimensional scenario for statistical inference. The methods provided in this package can help users to analyze high-dimensional time series data and make reliable predictions. To improve computational efficiency, the \pkg{HDTSA} package integrates \proglang{C++} through the \pkg{Rcpp} package. We illustrate the functions of the \pkg{HDTSA} package using simulated examples and real-world applications from finance and economics.

\end{abstract}





\section[]{Introduction} \label{sec:intro}

In the era of big data, the analysis of multivariate time series data is increasingly important in many areas. Examples include analyzing and predicting economic trends with a large number of macroeconomic variables \citep{DEMOL2008318}, portfolio selection and volatility matrix estimation based on large financial datasets \citep{Fan+Lv+Qi:2011}, weather forecasting with high-dimensional meteorological data \citep{Lam+Yao:2012}, and analyzing fMRI data \citep{Smith}. These applications require analyzing a $p$-dimensional vector time series or a $p \times q$ matrix time series, with the primary tasks typically divided into two categories: modeling and statistical inference. Hence, developing high-dimensional time series models and the associated statistical inference methods, especially when $p$ and $q$ are much larger than the sample size $n$, has become a recent focus in statistical research. 

On the theoretical side, various high-dimensional time series analysis methods have been developed in literature. Even when the dimensions $p$ and $q$ are moderately large, conventional vector time series models (such as vector autoregressive and moving average models) may face serious issues such as the lack of model identification and overparametrization, leading to difficulties in statistical inference. Dimension reduction methods have been proposed to reduce the number of parameters in high-dimensional vector time series models, including principal component analysis \citep[PCA,][]{Stock+Watson:2002,Wang+Han+Liu:2013, Chang+Guo+Yao:2018}, and factor models \citep{Bai,Lam+Yao+Bathia:2011, Lam+Yao:2012,Chang+Guo+Yao:2015}. For high-dimensional matrix time series, popular approaches to do dimension reduction include the Tucker decomposition \citep{Wang+Liu+Chen:2019, Chen+Chen:2022,Chen+yang+zhang:2022,Barigozzi+He:2023, Barigozzi:2023,Han+chen+yang+zhang:2024,lam2024} and the canonical polyadic (CP) decomposition \citep{Chang+He+Yang+Yao:2023,Chang+Du+Huang+Yao:2024,Chen+lam:2024,Han+Yang+Zhang+Chen:2024}.
Furthermore, cointegration also entails a dimension reduction for $p$-dimensional nonstationary vector time series. The early literature, for example \cite{Engle+Granger:1987} and \cite{Johansen:1991}, studied parametric settings such as the unit root vector autoregressive models. Recently, \cite{Zhang+Robinson+Yao:2019} proposed a model-free method for identifying the cointegration rank, which allows the integer-valued integration orders of the observed series to be unknown and different. In statistical inference, testing for white noise and martingale difference sequence are two fundamental problems. Classical white noise tests include the Box-Pierce test \citep{Box:1970}, the Ljung-Box test \citep{Ljung:1978}, the Hosking test \citep{Hosking:1980}, and the Li-McLeod test \citep{LiMcLeod:1981}.
These tests are shown to be asymptotically $\chi^2$-distributed based on the assumption that observations under the null hypothesis are independent and identically distributed (i.i.d.). Some studies focus on extending the asymptotic theory to accommodate white noise without i.i.d. assumption. One common approach is to establish the asymptotic normality of a standardized portmanteau test statistic, see, for example, \cite{Lobato:2001} and \cite{Shao:2011}. However, the convergence is typically slow.
\cite{LZ19} proposed a powerful portmanteau white noise test for high-dimensional time series. \cite{Chang+Zhou+Yao:2017}  proposed a new omnibus test for high-dimensional vector white noise which outperforms conventional methods. 
Most traditional martingale difference tests are limited to the univariate case. See \cite{Hong:1999} and \cite{Chen+Deo:2006} for examples. \cite{Hong+Linton+Zhang:2017} extended the univariate variance ratio test proposed by \cite{Chen+Deo:2006} to the multivariate scenario with $p \leq n$. However, this extension can only capture linear temporal dependence. More recently, \cite{Chang+Jiang+Shao:2022} proposed a novel martingale difference sequence test which is not only applicable in the high-dimensional scenario where $p$ can be much larger than $n$, but also capable of capturing nonlinear temporal dependence.

On the practical side, numerous tools for time series analysis are available in \proglang{R}. The traditional PCA can be implemented by the functions \code{prcomp()} and \code{princomp()} in the \pkg{stats} package. There are also many packages including functions to implement PCA, such as the packages \pkg{FactoMineR} \citep{FactoMineR} and \pkg{jvcoords} \citep{jvcoords}. However, such methods can not be used to analyze high-dimensional time series data, particularly when $p>n$. Traditional factor analysis can be implemented by the function \code{factanal()} in the \pkg{stats} package. For the high-dimensional scenario when $p>n$, the \code{apca()} function in the package \pkg{MTS} \citep{MTS} performs the asymptotic PCA introduced in Chapter 6 of \cite{MTSbook}; the \pkg{HDRFA} \citep{HDRFA} package provides two functions \code{HPCA()} and \code{HPCA_FN()} to do robust factor analysis based on the Huber loss, as well as the \code{PCA()} function to estimate the factor models by the method proposed by \cite{Bai}. As far as the matrix time series factor models are concerned, several of the aforementioned dimension reduction methods have been implemented in \proglang{R}, such as the packages \pkg{MEFM} \citep{MEFM}, \pkg{tensorTS} \citep{tensorTS}, \pkg{TensorPreAve} \citep{TensorPreAve}, and \pkg{RTFA} \citep{RTFA}. For the cointegration analysis, the traditional Engle-Granger test \citep{Engle+Granger:1987} can be implemented by the package \pkg{aTSA} \citep{aTSA}; and the traditional Johansen test \citep{Johansen:1991} can be implemented by the packages \pkg{urca} \citep{urca} and \pkg{tsDyn} \citep{tsDyn}. The package \pkg{cointReg} \citep{cointReg} provides algorithms for parameter estimation and statistical inference in cointegrating regression models. For the white noise test, the \code{Box.test()} function in the \pkg{stats} package is widely used for the univariate time series. For multivariate time series, the \pkg{MTS} package provides the function \code{mq()} to implement the Ljung-Box test, and the package \pkg{portes} \citep{portes} provides the functions \code{Hosking()}, \code{LjungBox()}, and \code{LiMcLeod()} for the Hosking test, the Ljung-Box test, and the Li-McLeod test, respectively. But these methods are not applicable for the high-dimensional scenario. Furthermore, as far as we are aware, there is no \proglang{R} package that implements the high-dimensional martingale difference tests. 

Despite the rapid development of time series analysis methods, to the best of our knowledge, there is no package for \proglang{R} or outside of the \proglang{R} environment that integrates different methods for high-dimensional time series analysis.

In this paper, we present the \pkg{HDTSA} package \citep{HDTSA:2024} for \proglang{R}, which is available from the Comprehensive \proglang{R} Archive Network (CRAN) at \url{https://CRAN. R-project.org/package=HDTSA}. It provides a general framework for analyzing high-dimensional time series. The \pkg{HDTSA} package includes four dimension reduction methods for high-dimensional time series modeling and two statistical inference methods for high-dimensional time series data. Specifically, for high-dimensional vector time series, the factor model \citep{Lam+Yao+Bathia:2011, Lam+Yao:2012, Chang+Guo+Yao:2015} and the PCA method \citep{Chang+Guo+Yao:2018} are integrated in this package. For high-dimensional matrix time series, the CP-decomposition method \citep{Chang+He+Yang+Yao:2023,Chang+Du+Huang+Yao:2024} in this package can be used. The package also provides a cointegration analysis method \citep{Zhang+Robinson+Yao:2019} for modeling nonstationary vector time series. For statistical inference, the \pkg{HDTSA} package provides a white noise test \citep{Chang+Zhou+Yao:2017} and a martingale difference test \citep{Chang+Jiang+Shao:2022} in the high-dimensional scenario. 
One of the main advantages of this package is that it incorporates recently developed methods that can be applied when the dimension of the vector (or matrix) time series is much larger than the sample size. These methods, with solid theoretical guarantees, have wide applications and outperform the conventional time series analysis methods in most high-dimensional scenarios. A second advantage of the \pkg{HDTSA} package is its integration with \proglang{C++}  through the \pkg{Rcpp} \citep{Rcpp:2024} interface, which significantly improves its computational efficiency, especially for high-dimensional matrix multiplications. Furthermore, the use of the \pkg{HDTSA} package is simple and flexible. The purpose of this paper is to provide guidance for the practical implementation of the \pkg{HDTSA} package and a review of the associated methods.

The paper is organized as follows. Section \ref{sec:model} provides a concise overview of the high-dimensional time series analysis methods used in the \pkg{HDTSA} package. Section \ref{sec:procedures} presents the main functions of the package, as well as illustrative examples. Section \ref{sec:realdata} demonstrates practical implementations of the package through two real-world data examples. Section \ref{sec:Concusion} gives conclusions and discussions of possible extensions. 

{\it Notation.}
For any positive integer $d \geq 2$,  write $[d] = \{1, \ldots, d\}$. For any $d$-dimensional vector $\bx = (x_1,\ldots,x_d)^{\T}$, we use $|\bx|_1 = \sum_{i=1}^d |x_i|$, $|\bx|_2 = (\sum_{i=1}^d x_i^2)^{1/2}$, and $|\bx|_\infty = \max_{i \in [d]} |x_i| $ to denote its $L_1$-norm, $L_2$-norm, and $L_{\infty}$-norm, respectively. Let $1(\cdot)$ denote the indicator function and $\bI_p$ denote the $p \times p$ identity matrix. For any real number $x$, let $\lceil x \rceil$ and $\lfloor x \rfloor$ denote the smallest integer not less than $x$ and the largest integer not greater than $x$, respectively. We use $\otimes$ to denote the Kronecker product operation. For any $m_1\times m_2$ matrix $\bH=(h_{i,j})$, let $\rm{vec}(\bH)$ denote the $(m_1m_2)$-dimensional vector obtained by stacking the columns of $\bH$. We use $\mathcal{N}(\boldsymbol{\mu},\boldsymbol{\Sigma})$ to denote the normal distribution with mean $\boldsymbol{\mu}$ and covariance matrix $\boldsymbol{\Sigma} $, and $\mathcal{U}(a,b)$ to denote a uniform distribution over the interval $[a,b]$. 

\section{Methodology} \label{sec:model}


\subsection{Factor model for vector time series}\label{sec:factors}

Let $\by_t$ be a $p$-dimensional time series. The factor model for $\by_t$ admits the form 
\begin{equation} \label{factor model}
	 \mathbf{y}_{t}=\mathbf{A} \mathbf{x}_{t}+\bepsilonb_{t}\,,
\end{equation}
where $\mathbf{x}_{t}$ is an $r$-dimensional latent factor process with unknown integer $r \le p$, $\mathbf{A}$ is a $p \times r$ unknown factor loading matrix, and $\bepsilonb_{t}$ is a $p$-dimensional vector white noise process. In this model, $\bA$ and $\mathbf{x}_{t}$ cannot be uniquely identified. Only the linear space spanned by the columns of $\bA$, denoted by $\mathcal{M}(\bA)$, can be uniquely determined. Define a $p\times p$ positive definite matrix
\begin{equation}\label{eq:bW}
	\mathbf{W}= \sum_{k=1}^{K}\mathbf{\Sigma}_y(k){\mathbf{\Sigma}_y(k)}^{\T}\,,
\end{equation}
where $K\geq1$ is a prescribed integer, and $$\mathbf{\Sigma}_y(k)=\frac{1}{n-k}\sum_{t=1}^{n-k}\mathrm{Cov}(\mathbf{y}_{t+k},\mathbf{y}_{t})\,.$$ 
As shown in \cite{Lam+Yao+Bathia:2011} and \cite{Lam+Yao:2012}, $r$ equals to the number of nonzero eigenvalues of $\bW$ and $\mathcal{M}(\mathbf{A})$ is identical to the linear space spanned by the eigenvectors of $\bW$ associated with its $r$ nonzero eigenvalues. Given the observations $\{\by_t\}_{t=1}^n$, let
\begin{equation}\label{eq:hatbW}
	\hat{\mathbf{W}}= \sum_{k=1}^{K}T_{\delta}\{\hat{\mathbf{\Sigma}}_y(k)\}T_{\delta}\{\hat{\mathbf{\Sigma}}_y(k)\}^{\T}~{\rm with}~\hat{\mathbf{\Sigma}}_{y}(k)=\frac{1}{n-k}\sum_{t=1}^{n-k} (\mathbf{y}_{t+k}-\bar{\mathbf{y}})(\mathbf{y}_{t}-\bar{\mathbf{y}})^{\T}\,,
\end{equation}
where $\bar{\mathbf{y}}=n^{-1}\sum_{t=1}^{n}\mathbf{y}_{t}$ and $T_{\delta}(\cdot)$ is a threshold operator defined as $T_{\delta}(\bW)=\{w_{i,j}1(|w_{i,j}|\geq \delta)\}$ for any matrix $\bW=(w_{i,j})$, with the threshold level $\delta \geq 0$. Here $\hat{\bW}$ specified in \eqref{eq:hatbW} provides an estimate for $\bW$. We can obtain the estimates of $r$ and $\mathcal{M}(\bA)$ via the eigenanalysis of $\hat{\bW}$. More specifically, let ${\lambda}_{1}(\hat{\bW}) \geq \dots \geq {\lambda}_{p}(\hat{\bW})\geq 0$ be  the eigenvalues of $\hat{\mathbf{W}}$. We can estimate $r$ as
\begin{equation}\label{thumb rule}
	\hat{r}=\arg \min_{i \in [R]}  \frac{{\lambda}_{i+1}(\hat{\bW})}{{\lambda}_{i}(\hat{\bW})}\,,
\end{equation}
where $R=\lfloor 0.75 p\rfloor$. Select $\hat{\bA}$ as a $p\times \hat{r}$ matrix with 
columns being the orthonormal eigenvectors associated with $\lambda_1(\hat{\bW}),\ldots,\lambda_{\hat{r}}(\hat{\bW})$. Then $\mathcal{M}(\hat{\bA})$ provides an estimate of $\mathcal{M}(\bA)$.
Such estimation procedure can also be extended to address the following two more complicated settings:

\begin{extension}[Including both strong and weak factors]\label{ext:strongweakFac}
	Consider model (\ref{factor model}) with $r_1$ strong factors and $r_2$ weak factors (i.e., $r=r_1+r_2$). See Remark 1 in  \cite{Lam+Yao:2012} for the definition of strong factors and weak factors. Then (\ref{factor model}) can be reformulated as 
	\begin{equation}
		\label{twostepmodel}
	\mathbf{y}_t=\mathbf{A}\mathbf{x}_t+\bepsilonb_t=\mathbf{A}_1\mathbf{x}_{t,1}+\mathbf{A}_2\mathbf{x}_{t,2}+\bepsilonb_t\,,
	\end{equation}
	where $\mathbf{x}_t=(\mathbf{x}_{t,1}^{\T}, \mathbf{x}_{t,2}^{\T} )^{\T}$, $\mathbf{A}=(\mathbf{A}_1,\mathbf{A}_2)$, $\bx_{t,1}$ consists of $r_1$ strong factors, and $\bx_{t,2}$ consists of $r_2$ weak factors. 
    \cite{Lam+Yao:2012} propose the following two-step method to estimate $\mathcal{M}(\bA)$: 
	\begin{leftbar}
    \noindent{\it Step 1.} Apply the ratio-based method \eqref{thumb rule} to $\hat{\mathbf{W}}$ given in \eqref{eq:hatbW}, and denote the associated estimate by $\hat{r}_1$. Select $\hat{\mathbf{A}}_1$ as a $p\times \hat{r}_1$ matrix with its columns being the orthonormal eigenvectors of $\hat{\mathbf{W}}$ associated with  its $\hat{r}_1$ largest eigenvalues. Let $\mathbf{y}_t^{*} = \mathbf{y}_t-\hat{\mathbf{A}}_1 {\hat{\mathbf{A}}_1}^{\T} \mathbf{y}_t$.
        
    \noindent{\it Step 2.} Define $\tilde{\bW}$ in the same manner as $\hat{\bW}$ with replacing $\by_t$ by $\by_t^*$. Apply the ratio-based method \eqref{thumb rule} to $\tilde{\bW}$ and denote the associated estimate by $\hat{r}_2$. Select $\hat{\bA}_2$ as a $p \times \hat{r}_2$ matrix with columns being the orthonormal eigenvectors of $\tilde{\bW}$ associated with its $\hat{r}_2$ largest eigenvalues. Then $\mathcal{M}(\hat{\bA})$ with 	%
    	$\hat{\mathbf{A}}=(\hat{\mathbf{A}}_1,\hat{\mathbf{A}}_2)$ provides an estimate for $\mathcal{M}(\bA)$.
	\end{leftbar}

\end{extension}

\begin{extension}[With observed regressors]
	\cite{Chang+Guo+Yao:2015} consider the regression model:
	\begin{equation} \label{regression}
		\mathbf{y}_t=\mathbf{D} \mathbf{z}_t + \mathbf{A} \mathbf{x}_t+ \bepsilonb_t\,,
	\end{equation}
	where $\mathbf{y}_t$ and $\mathbf{z}_t$ are, respectively, observable $p$-dimensional and $m$-dimensional time series, $\mathbf{x}_t$ is an $r$-dimensional latent factor process with zero mean and unknown integer $r\leq p$, $\bepsilonb_t $ is a $p$-dimensional vector white noise process, $\mathbf{D}$ is a $p\times m$ unknown regression coefficient matrix, and $\mathbf{A}$ is a $p\times r$ unknown factor loading matrix. 
 The regression coefficient matrix $\mathbf{D}$ can be estimated from the linear regression $\by_t=\bD \bz_t+\bfeta_t$, where $\bfeta_t=\bA\bx_t+\bepsilonb_t$ is treated as the error term. Given an estimate $\hat{\bD}$, we can define $\hat{\bW}_{\eta}$ in the same manner as $\hat{\bW}$ given in  \eqref{eq:hatbW} with replacing $\by_t$ by $\hat{\bfeta}_t=\mathbf{y}_t-\hat{\mathbf{D}} \mathbf{z}_t$, and then obtain the estimate of $\mathcal{M}(\bA)$  by applying the proposed eigenanalysis procedure to $\hat{\bW}_{\eta}$.
\end{extension}

\subsection{PCA for vector time series}\label{sec:pca}

Let $\by_t$ be an observable $p$-dimensional weakly stationary time series. The PCA for $\mathbf{y}_t$ considers the model
\begin{equation}\label{pca model}
	\mathbf{y}_t=\mathbf{A}\mathbf{x}_t~\textrm{with}~\mathbf{x}_t=\{\mathbf{x}_{t}^{(1),\T}, \dots, \mathbf{x}_{t}^{(q),\T}\}^{\T}\,,
\end{equation}
where $\mathbf{A}$ is a $p\times p$ unknown matrix, and $\mathbf{x}_t$ is an unobservable $p$-dimensional weakly stationary time series consisting of $q\ (>1)$ both contemporaneously and serially uncorrelated subseries, i.e., $\mathrm{Cov}\{\mathbf{x}_t^{(i)},\mathbf{x}_s^{(j)}\}=\mathbf{0}$ for all $t,s$ and $i \not= j$. If we can obtain such defined $(\bA,\bx_t)$, the modeling of $\by_t$ can be translated to model $q$ lower dimensional series $\mathbf{x}_t^{(1)},\dots,\mathbf{x}_t^{(q)}$ separately, which can avoid the challenges we encountered in modeling $\by_t$ directly when $p$ is large. 
As discussed below Equation (2.3) of \cite{Chang+Guo+Yao:2018}, we can assume ${\rm Var}(\by_t)=\bI_p={\rm Var}(\bx_t)$ without loss of generality. In practice, we can always replace $\by_t$ by $\hat{\bV}^{-1/2}\by_t$ in a preliminary step with $\hat{\bV}$ being a consistent estimator of ${\rm Var}(\by_t)$. Due to ${\rm Var}(\by_t)=\bI_p={\rm Var}(\bx_t)$, we know $\bA$ in \eqref{pca model} is an orthogonal matrix.

Let $p_j$ be the dimension of $\bx_t^{(j)}$ and write $\bA = (\bA_1,\ldots,\bA_q)$ with $\bA_j\in\mathbb{R}^{p\times p_j}$. Denote by $\mathcal{M}(\mathbf{A}_j)$ the linear space spanned by the columns of $\mathbf{A}_j$. For model \eqref{pca model}, only $\mathcal{M}(\mathbf{A}_1),\dots,\mathcal{M}(\mathbf{A}_q)$ can be uniquely identified.  Since $\bx_t^{(j)}=\bA_j^{\T}\by_t$ and $\bx_t^{(j)}$ is unobservable, we can select $\mathbf{x}_t^{(j)}$ as $\mathbf{\Gamma}_j^{\T}\by_t$ for any $\mathbf{\Gamma}_j\in\mathbb{R}^{p\times p_j}$ satisfying $\mathbf{\Gamma}_j^{\T}\mathbf{\Gamma}_j=\mathbf{I}_{p_j}$ and $\mathcal{M}(\mathbf{\Gamma}_j)=\mathcal{M}(\mathbf{A}_j)$. For $\bW$ given in \eqref{eq:bW}, we define
\begin{equation*}\label{eq:pca}
		\mathbf{W}_y=\bI_p+\bW\,.
\end{equation*}
As shown in \cite{Chang+Guo+Yao:2018}, the columns of $\bA$ can be selected as an appropriate permutation of the orthonormal eigenvectors of $\bW_y$. Let 
\begin{equation}\label{eq:threshWy}
	 \hat{\mathbf{W}}_y= \bI_p+\hat{\bW}
\end{equation}
with $\hat{\bW}$ specified in \eqref{eq:hatbW}, which provides an estimate of $\bW_y$. 
Perform eigenanalysis on $\hat{\bW}_y$ and obtain a $p\times p$ orthogonal matrix $\hat{\bgamma}$ with columns being the eigenvectors of $\hat{\bW}_y$. Let $\hat{\mathbf{z}}_t=\hat{\mathbf{\Gamma}}^{\T}\mathbf{y}_t=(\hat{z}_{1,t},\ldots,\hat{z}_{p,t})^{\T}$. Denote by $\rho_{i,j}(h)$ the cross correlation between the two component series $\hat{z}_{i,t}$ and $\hat{z}_{j,t}$ at lag $h$. We say $\hat{z}_{i,t}$ and $\hat{z}_{j,t}$ connected if
the null hypothesis
\begin{equation}\label{rho}
    H_{0,i,j}:\rho_{i,j}(h)=0\ \  \mathrm{for\ any\ }h=0,\pm1,\pm2,\ldots,\pm m
\end{equation}
is rejected, where $m\geq 1$ is a prescribed integer. \cite{Chang+Guo+Yao:2018} propose the following permutation procedure:
\begin{leftbar}
	\noindent{\it Step 1.} Start with $p$ groups with each group containing one component of $\hat{\bz}_t$ only.
	
	\noindent{\it Step 2.} Combine two groups together if one connected pair is split over the two groups.

    \noindent{\it Step 3.} Repeat Step 2 above until all connected pairs are within one group.
	\end{leftbar}

Then, we obtain $\hat{\bx}_t$ with its components being a permutation of the components of $\hat{\bz}_t$, which is segmented into $q$ uncorrelated subseries $\hat{\bx}_t^{(1)},\ldots,\hat{\bx}_t^{(q)}$. Since $\hat{\bx}_t = \hat{\bA}^{\T}\by_t$, $\hat{\bA}$ is then obtained by rearranging the order of the columns of $\hat{\mathbf{\Gamma}}$ accordingly. \cite{Chang+Guo+Yao:2018} suggest the following two methods for testing the null hypothesis \eqref{rho}. 
\begin{method}[Maximum cross correlation method]\label{mtd:acf} 
Let $\hat{\rho}_{i,j}(h)$ be the sample cross correlation between $\hat{z}_{i,t}$ and $\hat{z}_{j,t}$ at lag $h$. Define 
\begin{equation*}
    {L}_n(i,j)=\max\limits_{|h|\leq m}|\hat{\rho}_{i,j}(h)|\,.
\end{equation*}
Write $\aleph =p(p-1)/2$. We rearrange the $\aleph$ obtained ${L}_n(i,j)$'s in the descending order, i.e., ${L}_{(1)} \geq \cdots \geq {L}_{(\aleph)}$. Let
\begin{equation*}
    \hat{\ell}=\arg \max_{ j \in [R]} \frac{{L}_{(j)}}{{L}_{(j+1)}}\,,
\end{equation*}
where $R=\lfloor 0.75 \aleph\rfloor$. We reject $H_{0,i,j}$ for the pairs corresponding to ${L}_{(1)},\ldots,{L}_{(\hat{\ell})}$.
\end{method}

\begin{method}[FDR-based method]\label{mtd:FDR} For each pair $(i,j)$ and lag $h$, define $$p_{i,j}(h)=2\Phi\{-\sqrt{n}\hat{\rho}_{i,j}(h)\}\,,$$ where $\Phi(\cdot)$ denotes the distribution function of $\mathcal{N}(0,1)$. Let $p_{i,j}^{(1)}\leq \cdots \leq p_{i,j}^{(2m+1)}$ be the order statistics of $\{p_{i,j}(h):h=0,\pm 1,\ldots,\pm m\}$. The p-value for the null hypothesis $H_{0,i,j}$ specified in \eqref{rho} is given by pv$_{i,j}=\min_{l\in[2m+1]} p_{i,j}^{(l)}(2m+1)/l$. Arranging the $\aleph$ obtained p-values in the ascending order, i.e., ${\rm pv}_{(1)} \leq\cdots \leq {\rm pv}_{(\aleph)}$. For some small constant $\beta\in (0,1)$, let 
\begin{equation}\label{eq:FDR_beta}
    \hat{\ell}=\max\big\{k\in[\aleph]: {\rm pv}_{(k)}\leq k\beta/\aleph\big\}\,.
\end{equation}
We reject $H_{0,i,j}$ for the pairs corresponding to ${\rm pv}_{(1)},\ldots, {\rm pv}_{(\hat{\ell})}$. 
\end{method}

\subsection{CP-decomposition for matrix time series} \label{sec:CP}
Let $\bY_t = (y_{i,j,t})$ be a $p\times q$ matrix time series. The CP-decomposition for $\bY_t$ admits the form
\begin{align}\label{CP-Model}
    \bY_t = \bA \bX_t \bB^{\T} + \bepsilonb_t\,,
\end{align}
where $\bX_t={\rm diag}(x_{t,1},\dots,x_{t,d})$ is a $d\times d$ unobservable diagonal matrix, $\bepsilonb_t$ is a $p \times q$ matrix white noise, $\bA=(\ba_1,\dots,\ba_d)$ and $\bB = (\bb_1,\dots,\bb_d)$ are, respectively, $p\times d$ and $q \times d$ constant matrices with $|\ba_l|_2=|\bb_l|_2=1$ for each $l \in [d]$, and $1\leq d<\min(p, q)$ is an unknown fixed integer. 

When $\text{rank}(\mathbf{A}) = d=\text{rank}(\mathbf{B})$, \cite{Chang+He+Yang+Yao:2023} propose the following two estimation methods for $(d,\bA,\bB)$.

\begin{method}[A direct estimation for $\bA$, $\bB$, and $d$]\label{mtd:direct}
Without loss of generality, we assume $q\leq p$. Let $\xi_t$ be a linear combination of the $pq$ elements of $\bY_t$. For any $k \geq 1$, define
\begin{align*}
    \bSigma_{\bY,\xi}(k) =  \frac{1}{n-k} \sum_{t=k+1}^{n}\mathbb{E}\big[\{\bY_t-\mathbb{E}(\bar{\bY})\}\{\xi_{t-k}-\mathbb{E}(\bar{\xi})\}\big]\,, 
\end{align*}
where $\bar{\bY}=n^{-1}\sum_{i=1}^n \bY_t$ and $\bar{\xi}=n^{-1}\sum_{i=1}^n \xi_t$. Let $\bB^{+} \equiv (\bb^1,\ldots,\bb^d)^{\T}$ be the Moore-Penrose inverse of $\bB$. As shown in \cite{Chang+He+Yang+Yao:2023}, the rows of $\bB^{+}$ are the eigenvectors of the generalized eigenequation 
\begin{align}\label{GEV_K1qK2q}
   \bK_{2,q} \bb = \lambda \bK_{1,q} \bb
\end{align}
with $\bK_{1,q}=\bSigma_{\bY,\xi}(1)^{\T}\bSigma_{\bY,\xi}(1)$ and $\bK_{2,q}=\bSigma_{\bY,\xi}(1)^{\T} \bSigma_{\bY,\xi}(2)$. Given the observations $\{\bY_t\}_{t=1}^n$, let 
\begin{align}\label{sigmayk}
    \hat{\bSigma}_k = T_{\delta} \{ \hat{\bSigma}_{\bY,\xi}(k) \}~\textrm{with}~\hat{\bSigma}_{\bY,\xi}(k) =  \frac{1}{n-k} \sum_{t=k+1}^{n}(\bY_t-\bar{\bY})(\xi_{t-k}-\bar{\xi})\,,
\end{align}
where $T_{\delta}(\cdot)$ is a threshold operator with the threshold level $\delta\geq 0$. Here $\hat{\bK}_{1,q}=\hat{\bSigma}_1^{\T}\hat{\bSigma}_1$ and $\hat{\bK}_{2,q}=\hat{\bSigma}_1^{\T}\hat{\bSigma}_2$ provide the estimates of $\bK_{1,q}$ and $\bK_{2,q}$, respectively. Let ${\lambda}_1(\hat{\bK}_{1,q}) \geq \dots \geq {\lambda}_q(\hat{\bK}_{1,q}) \geq 0$ be the eigenvalues of the $q\times q$ matrix $\hat{\bK}_{1,q}$. We can estimate $d$ as
\begin{align}\label{hatd}
    \hat{d} = \arg\min_{j\in [R]}\frac{{\lambda}_{j+1}(\hat{\bK}_{1,q})}{{\lambda}_{j}(\hat{\bK}_{1,q})},
\end{align}
where $R=\lfloor 0.75 q\rfloor$. Applying the spectral decomposition to $\hat{\bK}_{1,q}$, we have $\hat{\bK}_{1,q}=\hat{\bgamma}\hat{\bf \Lambda}\hat{\bgamma}^{\T}$, where $\hat{\bgamma}=(\hat{\boldsymbol{\gamma}}_1,\dots,\hat{\boldsymbol{\gamma}}_q)$ is a $q\times q$ orthogonal matrix and $\hat{\bf \Lambda}={\rm diag}\{\lambda_1(\hat{\bK}_{1,q}),\dots,\lambda_q(\hat{\bK}_{1,q})\}$. For $\hat{d}$ defined in \eqref{hatd}, define
\begin{align}
     \widetilde{\bK}_{1,q}=\sum_{j=1}^{\hat{d}} \lambda_j(\hat{\bK}_{1,q})\hat{\boldsymbol{\gamma}}_j \hat{\boldsymbol{\gamma}}_j^{\T}\,,
\end{align}
 which is a truncated version of $\hat{\bK}_{1,q}$. 
Let $\hat{\bb}^1,\dots,\hat{\bb}^{\hat{d}}$ be the eigenvectors of the generalized eigenequation
$
    \hat{\bK}_{2,q}\bb = \lambda\tilde{\bK}_{1,q}\bb$, 
which is a sample version of \eqref{GEV_K1qK2q}. The columns of $\bA$ and $\bB$ can be estimated as follows:
\begin{align}
    \hat{\ba}_l=\frac{\hat{\bSigma}_1\hat{\bb}^{l}}{|\hat{\bSigma}_1\hat{\bb}^{l}|_2} ~\mbox{and}  ~ \hat{\bb}_l=\frac{\hat{\bSigma}_1^{\T}\hat{\ba}^{l}}{|\hat{\bSigma}_1^{\T}\hat{\ba}^{l}|_2}\,, \quad l\in[\hat{d}]\,,
\end{align}
where $\hat{\bA}^{+}=(\hat{\ba}^1,\dots,\hat{\ba}^{\hat{d}})^{\T}$ is the Moore-Penrose inverse of $\hat{\bA}=(\hat{\ba}_1,\dots,\hat{\ba}_{\hat{d}})$.
\end{method}

Solving the generalized eigenequation \eqref{GEV_K1qK2q} defined by rank-reduced matrices could be a complex computational task. Hence, \cite{Chang+He+Yang+Yao:2023} propose a refinement (Method \ref{mtd:refined}) that reduces the $q$-dimensional rank-reduced generalized eigenequation to a $d$-dimensional full-ranked one.

\begin{method}[A refined estimation for $\bA$, $\bB$, and $d$]\label{mtd:refined}
 Without loss of generality, we assume $q\leq p$. For a prescribed integer $K \geq 1$, define
\begin{align}\label{eq:M1M2}
    \hat{\bM}_1 = \sum_{k=1}^{K}\hat{\bf \Sigma}_k\hat{\bf \Sigma}_k^{\T}~\textrm{and}~\hat{\bM}_2 = \sum_{k=1}^{K}\hat{\bf \Sigma}_k^{\T}\hat{\bf \Sigma}_k
\end{align}
for $\hat{\bf \Sigma}_k$ specified in \eqref{sigmayk}. Let ${\lambda}_1(\hat{\bM}_1)\geq \dots \geq {\lambda}_p(\hat{\bM}_1) \geq 0$ be the eigenvalues of the $p\times p$ matrix $\hat{\bM}_1$. Analogous to \eqref{hatd}, we can also estimate $d$ as
\begin{align*}
    \hat{d} = \arg\min_{j\in [R]} \frac{{\lambda}_{j+1}(\hat{\bM}_{1})}{{\lambda}_{j}(\hat{\bM}_{1})}
\end{align*}
 for $R$ specified in \eqref{hatd}. 
Let $\hat{\bP}$ be the $p\times \hat{d}$ matrix of which the columns are the $\hat{d}$ orthonormal eigenvectors of $\hat{\bM}_1$ associated with its $\hat{d}$ largest eigenvalues, and $\hat{\bQ}$ be the $q\times \hat{d}$ matrix of which the columns are the $\hat{d}$ orthonormal eigenvectors of $\hat{\bM}_2$ associated with its $\hat{d}$ largest eigenvalues. 
Define
\begin{align*}
      \tilde{\bf \Sigma}_{\bY}(k)= \frac{1}{n-k}\sum_{t=k+1}^{n}(\bY_t - \bar{\bY}) \otimes {\rm vec}(\bY_{t-k}-\bar{\bY}) ~\mbox{and}~ \hat{\bf \Theta}=\bI_p \otimes \{ (\hat{\bQ} \otimes \hat{\bP})\bw \}\,,
\end{align*}
where $\bw \in \mathbb{R}^{\hat{d}^2}$ is a given constant vector  with bounded $L_2$-norm. Let
\begin{align}\label{eq:hatSigmaZk}
    \check{\bf \Sigma}(k)= \hat{\bP}^{\T} \hat{\bf \Theta}^{\T} T_{\delta} \{ \tilde{\bf \Sigma}_{\bY}(k) \}  \hat{\bQ}\,,
\end{align}
where $T_{\delta}(\cdot)$ is a threshold operator with the threshold level $\delta\geq 0$.
Write 
\begin{align*}
    \hat{\bJ}=\{ \check{\bf \Sigma}(1)^{\T}\check{\bf \Sigma}(1) \}^{-1}\check{\bf \Sigma}(1)^{\T}\check{\bf \Sigma}(2)
\end{align*}
and let $\hat{\bv}^1,\dots,\hat{\bv}^{\hat{d}}$ be the $\hat{d}$ eigenvectors of the $d\times d$ matrix $\hat{\bJ}$. Let $\hat{\bU}=(\hat{\bu}_1,\dots,\hat{\bu}_{\hat{d}})$ and $\hat{\bV}=(\hat{\bv}_1,\dots,\hat{\bv}_{\hat{d}})$ with
\begin{align*}
    \hat{\bu}_l = \frac{ \check{\bf \Sigma}(1)\hat{\bv}^{l} }{ |\check{\bf \Sigma}(1)\hat{\bv}^{l}|_2 }~\mbox{and}~\hat{\bv}_l = \frac{ \check{\bf \Sigma}(1)^{\T}\hat{\bu}^{l} }{ |\check{\bf \Sigma}(1)^{\T}\hat{\bu}^{l}|_2 }\,,
\end{align*}
where $(\hat{\bu}^{1},\dots,\hat{\bu}^{\hat{d}})^{\T}$ is the inverse of $\hat{\bU}$. Then the estimators for $\bA$ and $\bB$ are defined as 
\begin{align*}
    \hat{\bA}=\hat{\bP}\hat{\bU}~\mbox{and}~\hat{\bB}=\hat{\bQ}\hat{\bV}\,.
\end{align*}
\end{method}

The validity of Methods \ref{mtd:direct} and \ref{mtd:refined} depends critically on the assumption ${\rm rank}(\mathbf{A}) =d={\rm rank}(\mathbf{B})$. However, such assumption may not always be satisfied in practice. Let $\text{rank}(\mathbf{A}) = d_1$ and $\text{rank}(\mathbf{B}) = d_2$ with some unknown $d_1,d_2\leq d$. \cite{Chang+Du+Huang+Yao:2024} propose a general estimation procedure for $(d_1,d_2,d,\bA,\bB)$ without requiring $d_1=d=d_2$. See Section 4 of \cite{Chang+Du+Huang+Yao:2024} for details.

\subsection{Cointegration analysis} \label{sec:cointegration}

For a $p$-dimensional time series $\by_t$, write $\nabla^0 \by_t = \by_t$, $\nabla \by_t =  \by_t- \by_{t-1}$,  and $\nabla^d \by_t = \nabla(\nabla^{d-1} \by_t)$ for any integer $d\geq 2$. Denote by $y_{i,t}$ the $i$th element of $\by_t$. For some nonnegative integers $d_1,\ldots,d_p$, we say $\by_t$ is an $I(d_1,\ldots,d_p)$ process if $\nabla^{d_i} y_{i,t}$ is an $I(0)$ process for each $i\in[p]$. The cointegration analysis for the  $I(d_1,\ldots,d_p)$ process $\by_t$  considers the model
\begin{equation}\label{eq:cointegration}
	\by_t=\bA\bx_t\,,
\end{equation}
where $\bA$ is an unknown and invertible constant matrix, $\bx_t=(\bx_{t,1}^{\T},\bx_{t,2}^{\T})^{\T}$ is a latent $p$-dimensional process, $\bx_{t,2}$ is an $r$-dimensional $I(0)$ process, $\bx_{t,1}$ is an $I(c_1,\ldots,c_{p-r})$ process with each $c_i \in \{d_1,\ldots,d_p\}$, and no linear combination of $\bx_{t,1}$ is $I(0)$. Each element of $\bx_{t,2}$ is a cointegrating error of $\by_t$ and $r\geq 0$ is the cointegration rank. 

The pair $(\bA,\bx_t)$ in \eqref{eq:cointegration} is not uniquely identified. Without loss of generality, we assume $\bA$ is an orthogonal matrix. Write $\bA=(\bA_1,\bA_2)$, where $\bA_1$ and $\bA_2$ are, respectively, $p\times(p-r)$ and $p\times r$ matrices. The linear space spanned by the columns of $\bA_2$, denoted by $\mathcal{M}(\bA_2)$, is called the cointegration space which is uniquely identified.  For a prescribed integer $K\geq1$, let 
\begin{align}\label{eq:CointhatW}
	\check{\mathbf{W}}= \sum_{k=0}^{K}\hat{\mathbf{\Sigma}}_y(k)\hat{\mathbf{\Sigma}}_y(k)^{\T}\,,
\end{align}
for $\hat{\mathbf{\Sigma}}_y(k)$ specified in \eqref{eq:hatbW}. When $r$ is known, \cite{Zhang+Robinson+Yao:2019} show that $\mathcal{M}(\bA_2)$ can be estimated by the linear space spanned by the $r$ eigenvectors of $\check{\bW}$ associated with the $r$ smallest eigenvalues, and $\mathcal{M}(\bA_1)$ can be estimated by the linear space spanned by the $(p-r)$ eigenvectors of $\check{\bW}$ associated with the $(p-r)$ largest eigenvalues. Denote by $\hat{\bA}$ the $p\times p$ orthogonal matrix with columns being the eigenvectors of $\check{\bW}$ associated with the eigenvalues arranged in descending order. Write $\hat{\bx}_t=\hat{\bA}^{\T}\by_t\equiv (\hat{x}_t^1,\ldots,\hat{x}_t^p)^{\T}$. \cite{Zhang+Robinson+Yao:2019} suggest the following two methods to determine $r$. 
\begin{method} \label{cointegrationRank1}
	 For $i\in[p]$, let $S_i(m)=\sum_{k=1}^{m}\hat{\rho}_i(k)$, where 
    $$\hat{\rho}_i(k)=\frac{n\sum_{t=1}^{n-k}(\hat{x}_{t+k}^i-\bar{\hat{x}}^i)(\hat{x}_{t}^i-\bar{\hat{x}}^i)}{(n-k)\sum_{t=1}^{n}(\hat{x}_{t}^i-\bar{\hat{x}}^i)^2}~\textrm{with}~\bar{\hat{x}}^i=\frac{1}{n}\sum_{t=1}^{n}\hat{x}_t^i\,. $$
	Then $r$ can be estimated as  
	\begin{align}\label{eq:Coint_c0}
	    \hat{r}=\sum_{i=1}^{p}1\bigg\{\frac{S_i(m)}{m}<c_0 \bigg\}
	\end{align}
	for some constant $c_0\in(0,1)$ and some large constant $m$.
\end{method}

\begin{method}\label{cointegrationRank2}
	For $i \in [p]$, consider the null hypothesis 
    \begin{align*}
        H_{0,i}:\hat{x}_t^{p-i+1} \sim I(0)\,.
    \end{align*}
    The estimation procedure for $r$ can be implemented as follows:
 \begin{leftbar}
	\noindent{\it Step 1.} Start with $i=1$. Perform the unit root test proposed by \cite{Chang+Cheng+Yao:2021} for $H_{0,i}$.
    
	\noindent{\it Step 2.} If the null hypothesis is not rejected at the significance level $\alpha$, increment $i$ by 1 and repeat Step 1. Otherwise, stop the procedure and denote the value of $i$ at termination as $i_0$. The cointegration rank is then estimated as $\hat{r}=i_0-1$.
	\end{leftbar}
\end{method}


\subsection{Testing for white noise}\label{sec:white noise}

Let $\mathbf{y}_{t}$ be a $p$-dimensional weakly stationary time series with mean zero. Consider the white noise hypothesis testing problem:
\begin{equation}\label{eq:wntest}
	H_0:\mathbf{y}_{t}\ \mathrm{is\ white\ noise\ \ \ versus\ \ \ }H_1:\mathbf{y}_{t}\ \mathrm{is\ not\ white\ noise\,.}
 \end{equation}
Let
$
	\hat{\bgamma}(k)=\big[\mathrm{diag}{\{\hat{\mathbf{\Sigma}}_{y}(0)\}}\big]^{-1/2}\hat{\mathbf{\Sigma}}_{y}(k)\big[\mathrm{diag}{\{\hat{\mathbf{\Sigma}}_{y}(0)\}}\big]^{-1/2}\equiv\{\hat{\rho}_{i,j}(k)\}$
for $\hat{\mathbf{\Sigma}}_y(k)$ specified in \eqref{eq:hatbW}.  \cite{Chang+Zhou+Yao:2017} suggest the following test statistic for \eqref{eq:wntest}:
\begin{equation}\label{T_n1}
	T_{n,{\rm WN}}=\max \limits_{k\in[K]} \max \limits_{i,j\in[p]} n^{1/2}| \hat{\rho}_{i,j}(k)|\,,
\end{equation}
where $K\geq 1$ is a prescribed integer. Write $\tilde{n}=n-K$ and define
$$
\mathbf{f} _t=\big( [\mathrm{vec}\{(\mathbf{y}_{t+1}-\bar{\mathbf{y}})(\mathbf{y}_{t}-\bar{\mathbf{y}})^{\T}\} ]^{\T},\dots, [\mathrm{vec}\{(\mathbf{y}_{t+K}-\bar{\mathbf{y}})(\mathbf{y}_{t}-\bar{\mathbf{y}})^{\T}\}]^{\T}  \big)^{\T}\,.
$$
Let $\bTheta$ be an $\tilde{n} \times \tilde{n}$ matrix with $(i,j)$th element $\mathcal{K}\{(i-j)/b_n\}$, where $\mathcal{K}(\cdot)$ is a kernel function and $b_n$ is the bandwidth. In practice, we can use the Bartlett kernel, Parzen kernel and Quadratic spectral kernel as suggested by \cite{Andrews:1991}, and choose $b_n$ following the data-driven method introduced in Section 6 of \cite{Andrews:1991}. For the significance level $\alpha\in(0,1)$, \cite{Chang+Zhou+Yao:2017} propose the following bootstrap procedure to obtain the critical value for $T_{n,{\rm WN}}$:

\begin{leftbar}
	\noindent{\it Step 1.} Let $B$ be some prescribed large constant. For $i \in [B]$, generate $\tilde{n}$-dimensional random vector $\bfeta_i=(\eta_{i,1},\ldots,\eta_{i,\tilde{n}})^{\T}$ independently from $\mathcal{N}(\bzero,\bTheta)$, which is also independent of $\{\by_t\}_{t=1}^n$. Let $\mathbf{g}_i=\tilde{n}^{-1/2}\{({\bf I}_K \otimes \hat{{\bOmega}})( \sum_{t=1}^{\tilde{n}} \eta_{i,t}\mathbf{f}_t)\}$, where $\hat{\bOmega}=\big[\mathrm{diag}{\{\hat{\mathbf{\Sigma}}_{y}(0)\}}\big]^{-1/2}\otimes\big[\mathrm{diag}{\{\hat{\mathbf{\Sigma}}_{y}(0)\}}\big]^{-1/2}$. 
	
	\noindent{\it Step 2.} Take the $\lfloor B\alpha \rfloor$th largest value among $| \mathbf{g}_1 |_\infty,\dots,| \mathbf{g}_B |_\infty$ as the critical value $\hat{\mathrm{cv}}_{\alpha}$.
\end{leftbar}

Then we reject the null hypothesis $H_0$ specified in \eqref{eq:wntest} if $T_{n,{\rm WN}}>\hat{\mathrm{cv}}_{\alpha}$. Furthermore, the performance of the proposed test can be further enhanced by first applying time series PCA \citep{Chang+Guo+Yao:2018} introduced in Section \ref{sec:pca} to $\by_t$. Specifically, we can apply the proposed white noise test to the $p$-dimensional time series $\hat{\bx}_{t}$ specified in Section \ref{sec:pca}. 

\subsection{Testing for martingale difference}\label{sec:martingale}

Let $\by_t$ be a $p$-dimensional time series with mean zero, and denote by $\mathcal{F}_s$ the $\sigma$-field generated by $\{\by_t\}_{t\leq s}$. We call $\{\by_{t}\}_{t=1}^n$ a martingale difference sequence (MDS) if  $\mathbb{E}(\by_t\,|\,\mathcal{F}_{t-1})={\bf 0}$ for any $t \in \mathbb{Z}$. Consider the martingale difference hypothesis testing problem:
\begin{align}\label{eq:MDStest}
	H_0:\ \{\mathbf{y}_{t}\}_{t=1}^n\ \mathrm{is\ a\ MDS\ \ versus\ \ }H_1:\ \{\mathbf{y}_{t}\}_{t=1}^n\ \mathrm{is\ not\ a\ MDS}\,.
\end{align}
 For the user-specified map $\boldsymbol{\phi}:\mathbb{R}^p \to \mathbb{R}^d$, write $\hat{\bbeta}_k=(n-k)^{-1} \sum_{t=1}^{n-k} \mathrm{vec} \{ \boldsymbol{\phi}(\mathbf{y}_t) \mathbf{y}_{t+k}^{\T} \}$ for each $k\geq 1$. \cite{Chang+Jiang+Shao:2022} suggest the following test statistic for \eqref{eq:MDStest}:
\begin{equation}\label{T_n2}
	T_{n,{\rm MDS}} = n \sum_{k=1}^{K} | \hat{\bbeta}_k |_\infty^2\,,
\end{equation}
where $K\geq 1$ is a prescribed integer. Let $\tilde{n}=n-K$ and define 
$$
\mathbf{f}_t=\big([ \mathrm{vec}\{\boldsymbol{\phi}(\mathbf{y}_t) \mathbf{y}_{t+1}^{\T}\}]^{\T},\dots, [\mathrm{vec}\{ \boldsymbol{\phi}(\mathbf{y}_t)\mathbf{y}_{t+K}^{\T}\}]^{\T} \big)^{\T}\,.
$$
Write $\left[ pd \right]_k :=\{(k-1)pd+1,\dots , kpd\}$ for $k \in [K]$. For $\bTheta$ specified in Section \ref{sec:white noise}, \cite{Chang+Jiang+Shao:2022} propose the following bootstrap procedure to obtain the critical value for $T_{n,{\rm MDS}}$:

\begin{leftbar}
	\noindent{\it Step 1.} Let $B$ be some prescribed large constant. For $i \in [B]$, generate $\tilde{n}$-dimensional vector $\bfeta_i=(\eta_{i,1},\ldots,\eta_{i,\tilde{n}})^{\T}$ independently from $\mathcal{N}(\bzero,\bTheta)$, which is also independent of $\{\by_t\}_{t=1}^n$. Let $G_i=\sum_{k=1}^{K} \max_{j \in [ pd ]_k } |g_{i,j}|^2$ with $\mathbf{g}_i =(g_{i,1},\ldots,g_{i,Kpd})^{\T} =\tilde{n}^{-1/2} \sum_{t=1}^{\tilde{n}} \eta_{i,t} (\mathbf{f}_t-\bar{\mathbf{f}})$, where $\bar{\mathbf{f}}=\tilde{n}^{-1} \sum_{t=1}^{\tilde{n}} \mathbf{f}_t$. 
	
	\noindent{\it Step 2.}  Take the $\lfloor B\alpha \rfloor$th largest value among $G_1,\dots,G_B$  as the critical value $\hat{\mathrm{cv}}_{\alpha}$.
\end{leftbar}
Then we reject the null hypothesis $H_0$ specified in \eqref{eq:MDStest} if $T_{n,{\rm MDS}}>\hat{\mathrm{cv}}_{\alpha}$.

\begin{table}
	\centering
        \caption{\label{tab:function} Summary of main functions in the \pkg{HDTSA} package }
        \resizebox{1\textwidth}{!}{
                \def\arraystretch{1.3}
	\begin{tabular}{p{2.5cm}<{\centering}p{7.0cm}<{\raggedright}p{5.5cm}<{\raggedright}}\hline
 
		{\bf Function}                     & {\bf Arguments} & {\bf Main outputs}    \\ \hline
		\code{Factors()}             & \code{Y}, \code{lag.k}, 
 \code{thresh}, \code{delta}, \code{twostep}     & \code{factor\_num}, \code{loading.mat}, \code{X}        \\ \hline
		\code{HDSReg()}              & \code{Y}, \code{Z}, \code{D}, \code{lag.k},  \code{thresh}, \code{delta}, \code{twostep}   & \code{factor\_num}, \code{reg.coff.mat},   \code{loading.mat}, \code{X}\\ \hline
		\code{PCA\_TS()}            & \code{Y}, \code{lag.k}, \code{opt}, \code{permutation}, \code{thresh}, \code{delta}, \code{prewhiten},  \code{m}, \code{beta}, \code{control}   & \code{B}, \code{X}, \code{NoGroups}, \code{No\_of\_Members}, \code{Groups}    \\ \hline
            \code{CP\_MTS()}             & \code{Y}, \code{xi}, \code{Rank}, \code{lag.k}, \code{lag.ktilde}, \code{method}, \code{thresh1},  \code{thresh2},  \code{thresh3}, \code{delta1}, \code{delta2}, \code{delta3} &\code{A}, \code{B}, \code{f}, \code{Rank}
            \\ \hline
		\code{Coint()}& \code{Y}, \code{lag.k}, \code{type}, \code{c0}, \code{m}, \code{alpha} &\code{A}, \code{coint\_rank}  \\ \hline
		\code{WN\_test()}              &\code{Y}, \code{lag.k}, \code{B}, \code{kernel.type}, \code{pre}, \code{alpha}, \code{control.PCA}     & \code{statistic}, \code{p.value}       \\ \hline
		\code{MartG\_test()} &\code{Y}, \code{lag.k}, \code{B}, \code{type}, \code{alpha}, \code{kernel.type}      &\code{statistic}, \code{p.value}    \\ \hline
		
		
	\end{tabular}
}	
\end{table}

%
\section[]{Practical implementation}\label{sec:procedures}

In this section, we give an overview of how the main functions in the \pkg{HDTSA} package work by some simulated examples. Table~\ref{tab:function} presents a list of the main functions implemented in this package, as well as their associated arguments and the corresponding outputs.

\subsection[]{Factor model for vector time series in \proglang{R}}\label{sec:factors_fun}

The factor modeling methods described in Section \ref{sec:factors} are implemented in the \pkg{HDTSA} package through the functions \code{Factors()} and \code{HDSReg()}. 

The function \code{Factors()} is used to estimate the number of factors $r$ and the factor loading matrix $\bA$ in the factor model \eqref{factor model}. The function with its default arguments is:
\begin{Code}
Factors(Y, lag.k = 5, thresh = FALSE, delta = 2 * sqrt(log(ncol(Y)) / nrow(Y)),
       twostep = FALSE) 
\end{Code}
where: 
\begin{itemize}
	\item{
		\code{Y} is the $n \times p$ data matrix $\bY=(\by_1,\ldots,\by_n)^{\T}$, where $\{\by_t\}_{t=1}^n$ are the observations of $p$-dimensional times series.
	}
	\item{
		\code{lag.k} represents the time lag $K$ involved in \eqref{eq:bW} with the default value being 5.
	}
        \item{
      \code{thresh} is a boolean indicating whether the threshold level $\delta$ involved in \eqref{eq:hatbW} should be $0$. The default value is \code{FALSE}, which indicates that $\delta=0$. If \code{thresh = TRUE}, the argument \code{delta} is used to specify the threshold level $\delta$, with a default value of $2(n^{-1}\log p)^{1/2}$.}
    
	\item{
		\code{twostep} is a boolean indicating if the two-step estimation procedure specified in Extension \ref{ext:strongweakFac} should be used. The default value is \code{FALSE}, which indicates that the standard estimation procedure in Section \ref{sec:factors} will be implemented.
	}
\end{itemize}
The function \code{Factors()} returns an object of class \code{"factors"}, which consists of the following parts:
\begin{itemize}
	\item{
		\code{factor_num}, the estimated number of factors $\hat{r}$.
	}
	\item{
		\code{loading.mat}, the estimated factor loading matrix $\hat{\bA}$.
	}
        \item{
		\code{X}, the $n\times \hat{r}$ matrix $\hat{\bf X}=(\hat{\bf x}_1,\dots,\hat{\bf x}_n)^{\T}$ with $\hat{\bf x}_t = \hat{\bA}^{\T}\hat{\bf y}_t$.
	}
\end{itemize}

The implementation of the \code{Factors()} function involves large matrix computations. For example, when calculating the matrix $\hat{\bW}$ defined in \eqref{eq:hatbW}, we need to perform matrix multiplications between two $p \times p$ matrices for $K$ times. This process is computationally expensive, particularly when $n$ and $p$ are large. To address this issue, we use the matrix multiplication operation provided by the \pkg{RcppEigen} package \citep{RcppEigen:2013} instead of using the \proglang{R}'s built-in matrix multiplication, which significantly reduces the computational time. For instance, when $(n, p, K) = (1000, 500, 5)$, calculating $\hat{\bW}$ using the \pkg{RcppEigen} operations is 5 to 10 times faster than using $\proglang{R}$'s built-in implementations on a macOS platform with an M1 Pro CPU. Similarly, for other functions in the \pkg{HDTSA} package, we always use the \pkg{RcppEigen} package for handling large matrix multiplications.


The function \code{HDSReg()} is used to estimate the number of factors $r$, the factor loading matrix $\bA$, and the regression coefficient matrix $\bD$ in the model \eqref{regression}. The function with its default arguments is:
\begin{Code}
HDSReg(Y, Z, D = NULL, lag.k = 5, thresh = FALSE, 
       delta = 2 * sqrt(log(ncol(Y)) / nrow(Y)), twostep = FALSE)
\end{Code}
The arguments \code{Y}, \code{lag.k}, \code{thresh}, \code{delta}, and \code{twostep} are the same as those in the function \code{Factors()}. The argument \code{Z} is the $n \times m$ data matrix $\bZ=(\bz_1,\ldots,\bz_n)^{\T}$ consisting of the observed regressors. The optional argument \code{D} represents the regression coefficient matrix $\bD$ specified in \eqref{regression}. If \code{D = NULL} (the default), we will estimate $\bD$ by the least square method and create \code{reg.coff.mat} as a part of the output of this function. Otherwise, \code{D} can be given by the \proglang{R} users. In the output of \code{HDSReg()}, \code{X} represents the $n\times \hat{r}$ matrix $\hat{\bf X}=(\hat{\bf x}_1,\dots,\hat{\bf x}_n)^{\T}$ with $\hat{\mathbf{x}}_t=\hat{\mathbf{A}}^{\T}(\mathbf{y}_t-\hat{\mathbf{D}} \mathbf{z}_t)$.



We give two simulated examples to illustrate the usage of the functions \code{Factors()} and \code{HDSReg()}. 

\medskip

\noindent{\bf Example 1.} Consider the model \eqref{factor model} with $(n,p,r) = (400, 200, 3)$. The factor $\bx_t$ follows the VAR(1) model
\begin{align*}
    \bx_t = \left( \begin{matrix}0.6  & 0 & 0 \\
                    0 &  -0.5   &  0  \\
                    0   &   0   &  0.3  \end{matrix} \right)\bx_{t-1} + {\bf e}_t\,,
\end{align*}
where ${\bf e}_t \stackrel{{\rm i.i.d.}}{\sim} \mathcal{N}({\bf 0}, {\bf I}_3)$. In the model \eqref{factor model}, let $\boldsymbol{\varepsilon}_t \stackrel{{\rm i.i.d.}}{\sim} \mathcal{N}({\bf 0}, {\bf I}_p)$ and generate the elements of $\bf A$ independently from $\mathcal{U}(-1, 1)$. 


%
\begin{CodeChunk}
	\begin{CodeInput}
R> library("HDTSA")
R> set.seed(0)
R> {
+    n <- 400; p <- 200; r <- 3
+    X <- mat.or.vec(n, r)
+    x1 <- arima.sim(model=list(ar = c(0.6)), n = n)
+    x2 <- arima.sim(model=list(ar = c(-0.5)), n = n)
+    x3 <- arima.sim(model=list(ar = c(0.3)), n = n)
+    X <- t(cbind(x1, x2, x3))
+    A <- matrix(runif(p * r, -1, 1), ncol = r)
+    eps <- matrix(rnorm(n * p), p, n)
+    Y <- t(A %*% X + eps)
R> }
	\end{CodeInput}
\end{CodeChunk}
We can call the \code{Factors()} function with the argument \code{twostep = FALSE} to estimate $r$ and $\bA$ by the standard estimation procedure.
\begin{CodeChunk}
	\begin{CodeInput}
R> fac <- Factors(Y, lag.k = 5, twostep = FALSE)
R> print(fac)
	\end{CodeInput}
	\begin{CodeOutput}
	Factor analysis for vector time series

The estimated number of factors = 3
Time lag = 5
	\end{CodeOutput}
\end{CodeChunk}
The estimated number of factors and factor loading matrix can be extracted by the following codes:
\begin{CodeChunk}
	\begin{CodeInput}
R> r_hat <- fac$factor_num
R> load_hat <- fac$loading.mat
	\end{CodeInput}
\end{CodeChunk}
In the data generating process of  Example 1, let the third column of $\bA$ be divided by $p^{0.25}$, which indicates that the first two factors in $\bx_t$ are strong factors and the third factor is a weak factor.
\begin{CodeChunk}
	\begin{CodeInput}
R> A[, 3] <- A[, 3] / p^(0.25)
R> Y <- t(A %*% X + eps)
	\end{CodeInput}
\end{CodeChunk}
If we still use the standard estimation procedure by calling the function \code{Factors()} with the argument \code{twostep=FALSE}, we may get $\hat{r}=2$, which is not equal to the true value $r=3$.
\begin{CodeChunk}
	\begin{CodeInput}
R> fac_no2step <- Factors(Y, lag.k = 5, twostep = FALSE)
R> print(fac_no2step)
	\end{CodeInput}
	\begin{CodeOutput}
	Factor analysis for vector time series

The estimated number of factors = 2
Time lag = 5
	\end{CodeOutput}
\end{CodeChunk}
We can call the function \code{Factors()} with the argument \code{twostep=TRUE} to estimate $r$ using the two-step estimation method introduced in Extension \ref{ext:strongweakFac}. It can correctly identify that there are 2 strong factors (estimated in the first step) and 1 weak factor (estimated in the second step), with the total estimated number of factors $\hat{r}=3$, which is equal to the true value of $r$.
\begin{CodeChunk}
 	\begin{CodeInput}
R> fac_2step <- Factors(Y, lag.k = 5, twostep = TRUE)
R> print(fac_2step)
	\end{CodeInput}
 	\begin{CodeOutput}
	Factor analysis for vector time series

The estimated number of factors in the first step = 2
The estimated number of factors in the second step = 1
The estimated number of factors = 3
Time lag = 5
	\end{CodeOutput}
\end{CodeChunk}
\noindent{\bf Example 2.} Consider the model \eqref{regression} with $(n, p, r)=(400,200,3)$. The factor $\bx_t$ is generated as described in Example 1, and $\bz_t$ follows the VAR(1) model
\begin{align*}
    \bz_t = \left( \begin{matrix}0.625  & 0.125 \\
                    0.125   &  0.625  \end{matrix} \right)\bz_{t-1} + {\bf e}_t\,,
\end{align*}
where ${\bf e}_t \stackrel{{\rm i.i.d.}}{\sim} \mathcal{N}({\bf 0}, {\bf I}_2)$. In the model \eqref{regression}, let $\boldsymbol{\varepsilon}_t \stackrel{{\rm i.i.d.}}{\sim} \mathcal{N}({\bf 0}, {\bf I}_p)$ and generate the elements of $\bA$ and $\bD$ independently from $\mathcal{U}(-2,2)$. 

\begin{CodeChunk}
	\begin{CodeInput}
R> set.seed(0)
R> {
+   m <- 2
+   Z <- mat.or.vec(m, n)
+   S1 <- matrix(c(5/8, 1/8, 1/8, 5/8), 2, 2)
+   Z[, 1] <- c(rnorm(m))
+   for(i in c(2 : n))  Z[, i] <- S1 %*% Z[, i-1] + c(rnorm(m))
+   A <- matrix(runif(p * r, -2, 2), ncol = r)
+   D <- matrix(runif(p * m, -2, 2), ncol = m)
+   Y <- D %*% Z + A %*% X + eps
+   Y <- t(Y); Z <- t(Z)
R> }
	\end{CodeInput}
\end{CodeChunk}
%
If the regression coefficients are known, we can call the \code{HDSReg()} function with the arguments \code{D} specified by its true value $\bD$. 
%
\begin{CodeChunk}
	\begin{CodeInput}
R> fac_reg1 <- HDSReg(Y, Z, D, lag.k = 5)
R> print(fac_reg1)
	\end{CodeInput}
	\begin{CodeOutput}
	Factor analysis with observed regressors for vector time series

The estimated number of factors = 3
Time lag = 5
\end{CodeOutput}
\end{CodeChunk}
We can also call the \code{HDSReg()} function with the arguments \code{D} being unspecifed. Then $\bD$ will be estimated by the least square method. The estimated $\hat{\bD}$ can be extracted as follows:  
\begin{CodeChunk}
	\begin{CodeInput}
R> fac_reg2 <- HDSReg(Y, Z, lag.k = 5)
R> reg.coff.mat <- fac_reg2$reg.coff.mat
	\end{CodeInput}
\end{CodeChunk}

\subsection[]{PCA for vector time series in \proglang{R}}\label{sec:pca_fun}

The PCA method for vector time series introduced in Section \ref{sec:pca} is implemented in the \pkg{HDTSA} package through the function \code{PCA_TS()}. The function with its default arguments is:

%
\begin{Code}
PCA_TS(Y, lag.k = 5, opt = 1, permutation = c("max", "fdr"), thresh = FALSE,
       delta = 2 * sqrt(log(ncol(Y)) / nrow(Y)), prewhiten = TRUE, m = NULL,
       beta, control = list())
\end{Code}
where: 
\begin{itemize}
	\item{
		\code{Y} is the $n \times p$ data matrix $\bY=(\by_1,\ldots,\by_n)^{\T}$, where $\{\by_t\}_{t=1}^n$ are the observations of $p$-dimensional times series.
	}
	\item{
		\code{lag.k} represents the time lag $K$ involved in \eqref{eq:threshWy} with the default value being 5.
	}
	\item{
		\code{opt} is used to choose which method will be implemented to get a consistent estimate $\hat{\bV}$ (or $\hat{\bV}^{-1}$) for the covariance (precision) matrix of $\by_t$. If \code{opt = 1}, $\hat{\bV}$ will be defined as the sample covariance matrix. If \code{opt = 2}, the precision matrix $\hat{\bV}^{-1}$ will be calculated by using the function \code{clime()} of the \proglang{R} package \pkg{clime} \citep{clime:2022} with the arguments passed by \code{control}.
	}
    \begin{itemize}
      \item  {\code{control} is a list of arguments passed to the function \code{clime()}, which contains five elements: \code{nlambda}, the number of values for the program-generated \code{lambda}; \code{lambda.max} and \code{lambda.min} are the maximum and minimum values of the program-generated \code{lambda}; \code{standardize}, a boolean indicating whether the variables will be standardized to have mean zero and unit standard deviation; and \code{linsolver}, a boolean indicating whether the \code{primaldual} or \code{simplex} method should be employed. The default is \code{nlambda = 100}, \code{lambda.max = 0.8}, \code{lambda.min = 1e-4} ($n>p$) or \code{lambda.min = 1e-2} ($n<p$), \code{standardize = FALSE}, and \code{linsolver = "primaldual"}.}
    \end{itemize}
    \item{
    \code{permutation} represents the method for identifying the connected pair components of $\hat{\bf z}_t$ in the permutation procedure. If \code{permutation = "max"} (default), the maximum cross correlation method (Method \ref{mtd:acf}) will be used. If \code{permutation = "fdr"}, the FDR-based method (Method \ref{mtd:FDR}) will be used with the following argument:
    \begin{itemize}
        \item \code{beta}, the predetermined constant $\beta$ involved in \eqref{eq:FDR_beta}.
    \end{itemize}
    }
    \item{
      \code{thresh} is a boolean indicating whether the threshold level $\delta$ involved in \eqref{eq:threshWy} should be $0$. The default value is \code{FALSE}, which indicates that $\delta=0$. If \code{thresh = TRUE}, the argument \code{delta} is used to specify the threshold level $\delta$, with a default value of $2(n^{-1}\log p)^{1/2}$.
    }
    \item{
    \code{prewhiten} is a boolean indicating whether the prewhitening should be performed before implementing the permutation procedure. If \code{prewhiten = TRUE} (default), we prewhiten each component series $\hat{z}_{i,t}$ specified in Section \ref{sec:pca} by fitting a univariate AR model with the order between 0 and 5 determined by AIC. See detailed discussions in \cite{Chang+Guo+Yao:2018}.
    }
    \item{
	\code{m} is the number involved in the null hypothesis \eqref{rho} with the default value being 10.
    }
\end{itemize}

The function \code{PCA_TS()} returns an object of class \code{"tspca"}, which mainly consists of the following parts:
\begin{itemize}
        \item{
		\code{B}, the $p \times p$ transformation matrix $\hat{\bf B}=\hat{\bf \Gamma}^{\T}\hat{\bf V}^{-1/2}$, where $\hat{\bf \Gamma}$ is specified in Section \ref{sec:pca} and $\hat{\bV}^{-1}$ is the estimated precision matrix of $\by_t$.}
	\item{
		\code{X}, the $n \times p$ matrix $\hat{\bX} = (\hat{\bf x}_1,\dots,\hat{\bf x}_n)^{\T}$ with $\hat {\bf x}_t=\hat{\bf B}{\bf y}_t$.
	}
    \item{
		\code{NoGroups}, the number of groups.
	}
	\item{
		\code{No_of_Members}, the number of members in each group.
	}
	\item{
		\code{Groups}, the indices of the components of $\hat{\bf x}_t$ that are contained in each group.}
\end{itemize}

We give the following simulated example to illustrate the usage of the function \code{PCA_TS()}.

\medskip

\begin{figure}[htbp]
\centering
\includegraphics[width=13cm]{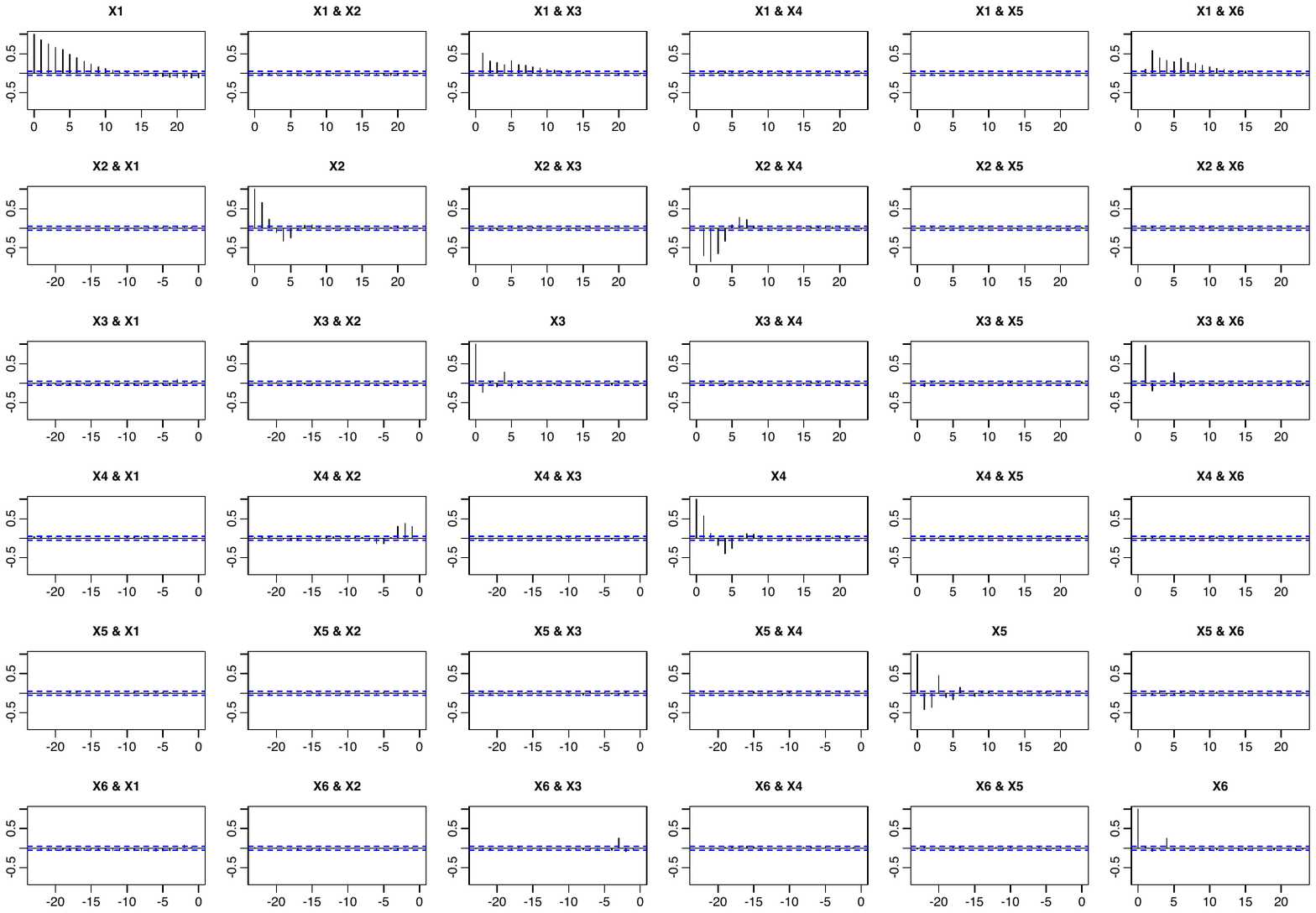}
\caption{\label{fig:acf} Cross correlogram of $\hat{\bx}_t = \hat{\bB}\by_t$ in Example 3.}
\end{figure}

\noindent{\bf Example 3.} Consider the model \eqref{pca model} with  $(n, p) = (1500, 6)$. In the model \eqref{pca model}, generate the elements of $\bf A$ independently from $\mathcal{U}(-3, 3)$. Let ${\bf x}_t = (x_{t,1},\ldots, x_{t,6})^{\T}$, where $x_{j,t}=\eta^{(1)}_{t+j-1}$ $(j=1,2,3)$, $x_{j,t}=\eta^{(2)}_{t+j-4}$ $(j=4,5)$ and $x_{6,t}=\eta^{(3)}_{t}$, where $\eta^{(1)}_{t}, \eta^{(2)}_{t}$ and $\eta^{(3)}_{t}$ follow the ARMA models:
\begin{align*}
    \eta^{(1)}_{t} & = 0.5 \eta^{(1)}_{t-1}+0.3\eta^{(1)}_{t-2} + \varepsilon^{(1)}_t - 0.9\varepsilon^{(1)}_{t-1} + 0.3\varepsilon^{(1)}_{t-2} + 1.2 \varepsilon^{(1)}_{t-3} + 1.3 \varepsilon^{(1)}_{t-4}\,, \\ 
    \eta^{(2)}_{t} & = 0.8 \eta^{(2)}_{t-1}-0.5\eta^{(2)}_{t-2} + \varepsilon^{(2)}_t + \varepsilon^{(2)}_{t-1} + 0.8\varepsilon^{(2)}_{t-2} + 1.8 \varepsilon^{(2)}_{t-3}\,, \\
    \eta^{(3)}_{t} & = - 0.7 \eta^{(3)}_{t-1}-0.5\eta^{(3)}_{t-2} + \varepsilon^{(3)}_t - \varepsilon^{(3)}_{t-1} - 0.8\varepsilon^{(3)}_{t-2}\,,
\end{align*}
and $\varepsilon^{(k)}_{t} \stackrel{{\rm i.i.d.}}{\sim} \mathcal{N}(0,1)$ $(k=1,2,3)$. 
Hence, ${\bf x}_t$ consists of three independent subseries with, respectively, 3, 2 and 1 components. 
%
\begin{CodeChunk}
	\begin{CodeInput}
R> library("HDTSA")
R> set.seed(0)
R> {
+   p <- 6; n <- 1500
+   X <- mat.or.vec(p, n)
+   x <- arima.sim(model=list(ar = c(0.5, 0.3), ma = c(-0.9, 0.3, 1.2,1.3)),
+                  n = n + 2, sd = 1)
+   for(i in 1:3) X[i,] <- x[i:(n + i - 1)]
+   x <- arima.sim(model=list(ar = c(0.8,-0.5), ma = c(1,0.8,1.8)),
+                  n = n + 1, sd = 1)
+   for(i in 4:5) X[i, ] <- x[(i - 3):(n + i - 4)]
+   x <- arima.sim(model=list(ar = c(-0.7, -0.5), ma = c(-1, -0.8)),
+                  n = n, sd = 1)
+   X[6,] <- x
+   A <- matrix(runif(p * p, -3, 3), ncol = p)
+   Y <- A %*% X
+   Y <- t(Y)
R> }
\end{CodeInput}
\end{CodeChunk}
To implement the permutation procedure by using the maximum cross correlation method, we call the \code{PCA_TS()} function with the argument \code{permutation = "max"}, while other arguments are set to their default values.
\begin{CodeChunk}
\begin{CodeInput}
R> result <- PCA_TS(Y, lag.k = 5, permutation = "max")
R> print(result)
\end{CodeInput}
\begin{CodeOutput}
	Principal component analysis for time series

Maximum cross correlation method
The number of groups = 3
The numbers of members in groups containing at least two members: 3 2
\end{CodeOutput}
\end{CodeChunk}
%
 The following result shows that the six components of $\hat{\bx}_t$ can be divided into 3 groups: $\{1, 3, 6\}, \{2, 4\}$ and $\{5\}$. 
\begin{CodeChunk}
\begin{CodeInput}
R> groups <- result$Groups
R> print(groups)
\end{CodeInput}
\begin{CodeOutput}
     Group 1 Group 2 Group 3
[1,]       1       2       5
[2,]       3       4       0
[3,]       6       0       0
\end{CodeOutput}
\end{CodeChunk}
The estimated $\hat{\bf x}_1,\ldots,\hat{\bf x}_n$ can be extracted from the \code{result} as follows:
\begin{CodeChunk}
	\begin{CodeInput}
R> X <- result$X
	\end{CodeInput}
\end{CodeChunk}
We can plot the cross correlogram of $\hat{\bx}_t$ by using the function \code{acf()} in the \pkg{stats} package, which is shown in Figure~\ref{fig:acf}. It can be seen from Figure~\ref{fig:acf} that the six components of $\hat{\bx}_t$ can be divided into three distinct groups: $\{1, 3, 6\}$, $\{2, 4\}$, and $\{5\}$, since there are no significant correlations across these three groups.
\begin{CodeChunk}
	\begin{CodeInput}
R> X=data.frame(X)
R> names(X) <- c("X1", "X2", "X3", "X4", "X5", "X6")
R> acf(X, ylab = " ", xlab = " ", plot = TRUE)
	\end{CodeInput}
\end{CodeChunk}

The permutation procedure by using the FDR-based method can be implemented by calling the \code{PCA_TS()} function with the argument \code{permutation = "fdr"}. 
\begin{CodeChunk}
\begin{CodeInput}
R> result <- PCA_TS(Y, lag.k = 5, permutation = "fdr", beta = 10^(-5))
\end{CodeInput}
\end{CodeChunk}

\subsection[]{CP-decomposition for matrix time series in \proglang{R}}\label{sec:cp_fun}

The CP-decomposition methods for matrix time series introduced in Section \ref{sec:CP} are implemented in the \pkg{HDTSA} package through the function \code{CP_MTS()}. The function with its default arguments is: 

\begin{Code}
CP_MTS(Y, xi = NULL, Rank = NULL, lag.k = 20, lag.ktilde = 10,
       method = c("CP.Direct", "CP.Refined", "CP.Unified"),
       thresh1 = FALSE, thresh2 = FALSE, thresh3 = FALSE,
       delta1 = 2 * sqrt(log(dim(Y)[2] * dim(Y)[3]) / dim(Y)[1]),
       delta2 = delta1, delta3 = delta1)
\end{Code}
where
\begin{itemize}
    \item{
    \code{Y} is an \(n \times p \times q\) data array, where its $t$-th slice along the first dimension is $\bY_t$, where $\{\bY_t\}_{t=1}^n$ are the observations of the $p \times q$ matrix time series.
    }
    \item{
    \code{xi} is an \(n \times 1\) vector $\boldsymbol{\xi} = (\xi_1,\ldots, \xi_n)^{\T}$, where $\xi_t$ represents a linear combination of $\bY_t$. If \code{xi = NULL} (the default), $\xi_{t}$ is determined by the PCA method introduced in Section 5.1 of \cite{Chang+He+Yang+Yao:2023}. Otherwise, \code{xi} can be given by the \proglang{R} users.
    }
    \item{
    \code{Rank} represents a list specifying the ranks \((d,d_1,d_2)\) specified in Section \ref{sec:CP}. The default value is \code{Rank = NULL}, which indicates that $d$, $d_1$ and $d_2$ are required to be estimated by the methods introduced in Section \ref{sec:CP}. Otherwise, they can be given by the \proglang{R} users.
    }

    \item{
    \code{lag.k}
    represents the time lag $K$ involved in \eqref{eq:M1M2}, and is used when \code{method} \code{= "CP.Refined"} or \code{method = "CP.Unified"}. The default value is 20.
    }
    \item{
    \code{lag.ktilde} represents the time lag $\tilde{K}$ involved in the unified estimation method, which is used when \code{method = "CP.Unified"}. Details can be found in Section 4.2 of \cite{Chang+Du+Huang+Yao:2024}. The default value is 10.
    }
    \item{
    \code{method} represents a character string specifying which CP-decomposition method to be used. It may take values \code{"CP.Direct"} (the direct estimation method introduced in Method \ref{mtd:direct}), \code{"CP.Refined"} (the refined estimation method introduced in Method \ref{mtd:refined}), or \code{"CP.Unified"} (the unified estimation method introduced in Section 4 of \cite{Chang+Du+Huang+Yao:2024}).
    } 
    \item{
      \code{thresh1} is a boolean indicating whether the threshold level $\delta$ involved in \eqref{sigmayk} should be $0$, which is used for all three methods. The default value is \code{FALSE}, which indicates that $\delta=0$. If \code{thresh1 = TRUE}, the argument \code{delta1} is used to specify the threshold level $\delta$, with a default value of $2\{n^{-1}\log (pq)\}^{1/2}$.
    }
    \item{
      \code{thresh2} is a boolean indicating whether the threshold level $\delta$ involved in \eqref{eq:hatSigmaZk} should be $0$, which is used only when \code{method = "CP.Refined"}. The default value is \code{FALSE}, which indicates that $\delta=0$. If \code{thresh2 = TRUE}, the argument \code{delta2} is used to specify the threshold level $\delta$, with a default value of $2\{n^{-1}\log (pq)\}^{1/2}$.
    }
    \item{
      \code{thresh3} is a boolean indicating whether the threshold level $\delta$ involved in (16) in Section 4.2 of \cite{Chang+Du+Huang+Yao:2024} should be $0$, which is used only when \code{method = "CP.Unified"}. The default value is \code{FALSE}, which indicates that $\delta=0$. If \code{thresh3 = TRUE}, the argument \code{delta3} is used to specify the threshold level $\delta$, with a default value of $2\{n^{-1}\log (pq)\}^{1/2}$.
    }
\end{itemize}
The output of \code{CP_MTS()} is an object of class \code{"mtscp"}, which consists the following parts:
\begin{itemize}
    \item{
    \code{A}, the estimated \(p \times \hat{d}\) left loading matrix \(\hat{\mathbf{A}}\).
    }
    \item{
    \code{B}, the estimated \(q \times \hat{d}\) right loading matrix \(\hat{\mathbf{B}}\).
    }
    \item{
    \code{f}, the estimated latent processes \(\hat{x}_{t,1},\ldots,\hat{x}_{t,\hat{d}}\).
    }
    \item{
    \code{Rank}, the estimated ranks $\hat{d}_1, \hat{d}_2$ and $\hat{d}$.
    }
\end{itemize}

We use the following simulated example to illustrate the usage of the function \code{CP_MTS()}.

\medskip

\noindent{\bf Example 4.} Consider the model \eqref{CP-Model} with $(n,p,q)=(400,10,10)$ and $d_1=d_2=d=3$. Let $\bA^\dagger \equiv (a^\dagger_{i,j})_{p \times d}$ and $\bB^\dagger \equiv (b^\dagger_{i,j})_{q \times d}$ with  $a^\dagger_{i,j}, b^\dagger_{i,j} \stackrel{{\rm i.i.d.}}{\sim} \mathcal{U}(-3,3)$
satisfying ${\rm rank}(\bA^\dagger) =d = {\rm rank}(\bB^\dagger)$. Define $\bP \in \mathbb{R}^{p\times d_1}$ and $\bQ\in\mathbb{R}^{q\times d_2}$ such that the columns of $\bP$ and $\bQ$ are, respectively, the $d_1$ and $d_2$ left-singular vectors corresponding to the $d_1$ and $d_2$ largest singular values of $\bA^{\dagger}$ and $\bB^{\dagger}$. Let $\bU^* = \bP^{\T} \bA^\dagger =(\bu_1^*,\ldots,\bu_d^*)$ and $\bV^* = \bQ^{\T} \bB^\dagger =(\bv_1^*,\ldots,\bv_d^* )$. 
Derive $\bU = (\bu_1,\ldots,\bu_d)$ and $\bV = (\bv_1,\ldots,\bv_d)$ with $\bu_j = \bu_j^*/ |\bu_j^*|_2$ and $\bv_j = \bv_j^*/ | \bv_j^* |_2$ for any $j \in [d]$. Write $\bx^*_j = (x^*_{1,j},\ldots,x^*_{n,j})^{\T}$ and let  $\bx^*_1,\ldots,\bx^*_d$ be $d$ independent AR(1) processes with independent $\mathcal{N}(0,1)$ innovations, and the autoregressive coefficients drawn from the uniform distribution on $[-0.95,-0.6]\cup [0.6,0.95]$. Let $\bX_t = \text{diag}( x_{t,1},\ldots,  x_{t,d} )$ with $ x_{t,j} = | \bv_j^* |_2 | \bu_j^* |_2  x^*_{t,j}$, $\bA=\bP\bU$ and $\bB=\bQ\bV$.
We generate $\bY_t = \bA \bX_t \bB^{\T} + \boldsymbol{\varepsilon}_t$, where $\boldsymbol{\varepsilon}_t = (\varepsilon_{t,i,j})_{p \times q}$ with  $\varepsilon_{t,i,j} \stackrel{{\rm i.i.d.}}{\sim} \mathcal{N}(0,1)$. 

This data generating process can be implemented using the function \code{DGP.CP()} in the \pkg{HDSTA} package. See the manual and the help file of the \code{DGP.CP()} function for usage details.


\begin{CodeChunk}
    \begin{CodeInput}
R> library("HDTSA")
R> p <- 10
R> q <- 10
R> n <- 400
R> d = d1 = d2 <- 3
R> set.seed(0)
R> data <- DGP.CP(n, p, q, d, d1, d2)
R> Y <- data$Y
    \end{CodeInput}
\end{CodeChunk}
We can use the direct estimation method for $\bA$, $\bB$ and $d$ by calling the function \code{CP_MTS()} with the argument \code{method = "CP.Direct"}. Then, we get $\hat{d}=3$.
\begin{CodeChunk}
    \begin{CodeInput}
R> res1 <- CP_MTS(Y, method = "CP.Direct")
R> print(res1)
\end{CodeInput}
\begin{CodeOutput}
	Estimation of matrix CP-factor model

Method: CP.Direct
The estimated number of factors d = 3
\end{CodeOutput}
Furthermore, the estimated loading matrices $\hat{\bA}$, $\hat{\bB}$ and the estimated factors $\hat{x}_{t,1},\ldots,\hat{x}_{t,d}$ can be extracted, respectively, by the following codes: 
\begin{CodeChunk}
    \begin{CodeInput}
R> A <- res1$A 
R> B <- res1$B 
R> f <- res1$f 
\end{CodeInput}
\end{CodeChunk}
The refined estimation method for $\bA$, $\bB$ and $d$ can be implemented by calling the function \code{CP_MTS()} with the argument \code{method = "CP.Refined"}. We also obtain $\hat{d} = 3$ by this method.
\begin{CodeInput}
R> res2 <- CP_MTS(Y, method = "CP.Refined")
R> print(res2)
\end{CodeInput}
\begin{CodeOutput}
	Estimation of matrix CP-factor model

Method: CP.Refined
The estimated number of factors d = 3
\end{CodeOutput}
We can also apply the unified estimation method by setting the argument \code{method = "CP.Unified"}. This function will return the estimates for $d_1, d_2$ and $d$ simultaneously. We get $\hat{d}_1=\hat{d}_2 = \hat{d} = 3$, which correctly identifies the true ranks.
\begin{CodeInput}
R> res3 <- CP_MTS(Y, method = "CP.Unified")
R> print(res3)
\end{CodeInput}
\begin{CodeOutput}
	Estimation of matrix CP-factor model

Method: CP.Unified
The estimated number of factors d = 3
The estimated rank of the left loading matrix d1 = 3
The estimated rank of the right loading matrix d2 = 3
\end{CodeOutput}
\end{CodeChunk}
For the cases where $d_1 < d$ or $d_2 < d$, only the unified estimation method is applicable. Consider Example 4 with $d_1=d_2=2$ and $d=3$ and generate data using the function \code{DGP.CP()}. Then, we get $\hat{d}_1=\hat{d}_2 = 2$ and $\hat{d} = 3$ by the unified estimation method.
\begin{CodeChunk}
    \begin{CodeInput}
R> d <- 3
R> d1 <- 2
R> d2 <- 2
R> set.seed(0)
R> data <- DGP.CP(n, p, q, d, d1, d2)
R> Y1 <- data$Y
R> res4 <- CP_MTS(Y1, method = "CP.Unified")
R> print(res4)
\end{CodeInput}
\begin{CodeOutput}
	Estimation of matrix CP-factor model

Method: CP.Unified
The estimated number of factors d = 3
The estimated rank of the left loading matrix d1 = 2
The estimated rank of the right loading matrix d2 = 2
\end{CodeOutput}
\end{CodeChunk}
Therefore, when $d_1$, $d_2$ and $d$ are unknown, we recommend using the unified estimation method. When $d_1$, $d_2$ and $d$ are known, we can call the function \code{CP_MTS()} by specifying them through the argument \code{Rank}.
\begin{CodeChunk}
    \begin{CodeInput}
R> res5 <- CP_MTS(Y1, Rank=list(d=3, d1=2, d2=2), method = "CP.Unified")
    \end{CodeInput}
\end{CodeChunk}


\subsection[]{Cointegration analysis in \proglang{R}}\label{sec:coint_fun}

The cointegration analysis method for vector time series introduced in Section \ref{sec:cointegration} is  implemented in the \pkg{HDTSA}
package through the function \code{Coint()}. The function with its default arguments is:
\begin{Code}
Coint(Y, lag.k = 5, type=c("acf", "urtest", "both"), c0 = 0.3, m = 20,
      alpha = 0.01)
\end{Code}
where

\begin{itemize}
        \item{
        \code{Y} is an \(n \times p \) data matrix $\bY = (\by_1,\dots,\by_n)^{\T}$, where $\{\by_t\}_{t=1}^n$ are the observations of $p$-dimensional times series.
        }
        \item{
        \code{lag.k} represents the time lag $K$ involved in \eqref{eq:CointhatW} with the default value being 5.
        }
        \item{
        \code{type} specifies the method used to identify the cointegration rank. It can take the following values: \code{"acf"} (the default) for the method introduced in Method \ref{cointegrationRank1}, \code{"urtest"} for the method introduced in Method \ref{cointegrationRank2}, and \code{"both"} to apply these two methods.}
        
        \item{
        \code{c0} represents the constant $c_0$ involved in \eqref{eq:Coint_c0}, which is used in Method \ref{cointegrationRank1} when \code{type = "acf"} or \code{type = "both"}. The default value is 0.3, as suggested in Section 2.3 of \cite{Zhang+Robinson+Yao:2019}.
        }
    
        \item{
        \code{m} represents the constant $m$ involved in \eqref{eq:Coint_c0}, which is used in Method \ref{cointegrationRank1} when \code{type = "acf"} or \code{type = "both"}. The default value is 20, as suggested in Section 2.3 of \cite{Zhang+Robinson+Yao:2019}.
        }
        \item{
        \code{alpha} represents the significance level $\alpha$ of the unit root tests, which is used in Method \ref{cointegrationRank2} when \code{type = "urtest"} or \code{type = "both"}. The default value is 0.01.
        }

\end{itemize}
The function \code{Coint()} returns an object of class \code{"coint"}, which consists of the following parts:
\begin{itemize}
	\item{
		\code{A}, the estimated \(\hat{\bA}\) described in Section \ref{sec:cointegration}.
	}
	\item{
		\code{coint_rank} represents the estimated cointegration rank $\hat{r}$.
	}
\end{itemize}

We use the following simulated example to illustrate the usage of the function \code{Coint()}. 

\medskip

\noindent{\bf Example 5.}  Consider the model \eqref{eq:cointegration} with $(n,p)=(1500,8)$. Let the first three components of ${\bf y}_t$ be 
\begin{gather*}
\begin{pmatrix}
y_{t,1} \\ y_{t,2}  \\ y_{t,3} 
\end{pmatrix}
=
\begin{pmatrix}
	1 & 1& 0\\ 0.5 & 0 & 1 \\ 0 &1 &0
\end{pmatrix}
\begin{pmatrix}
	x_{t,1} \\ x_{t,2} \\ x_{t,3}
\end{pmatrix}
=:{\bf A}_{11}
\begin{pmatrix}
x_{t,1} \\ x_{t,2} \\ x_{t,3}
\end{pmatrix}\,,
\end{gather*}
where $x_{t,1}$ is an $I(1)$ process  with independent $\mathcal{N}(0, 1)$ innovations, and $x_{t,2}, x_{t,3} \stackrel{{\rm i.i.d.}}{\sim} \mathcal{N}(0, 1)$. Let $y_{t,4}$ be a stationary ${\rm AR}(1)$ process, and $y_{t,5}, y_{t,6}, y_{t,7}, y_{t,8}$ be four ${\rm ARIMA}(1,1,1)$ processes. All the coefficients in ${\rm AR}(1)$ are $0.5$, the coefficients in ${\rm ARIMA}(1,1,1)$ are $(0.6, 0.8)$, and all the innovations are independent $\mathcal{N}(0, 1)$. Except for the elements in $\bA_{11}$ specified above, all the other elements of $\bA$ are generated independently from $\mathcal{U}(-3, 3)$. In this setting, we have the cointegration rank $r=3$. 
\begin{CodeChunk}
\begin{CodeInput}
R> library("HDTSA")
R> set.seed(0)
R> {
+   p <- 8; n <- 1500; r <- 3; d <- 1
+   X <- mat.or.vec(p, n)
+   X[1, ] <- arima.sim(n = n - d, model = list(order=c(0, d, 0)))
+   for(i in 2:3)X[i, ] <- rnorm(n)
+   for(i in 4:(r + 1)) X[i, ] <- arima.sim(model = list(ar = 0.5), n)
+   for(i in (r + 2):p) X[i, ] <- arima.sim(n = n - d,
+                                           model = list(order=c(1, d, 1),
+                                                        ar = 0.6, ma = 0.8))
+   A11 <- matrix(c(1, 1, 0, 1/2, 0, 1, 0, 1, 0), ncol = 3, byrow = TRUE)
+   A <- matrix(runif(p * p, -3, 3), ncol = p)
+   A[1:3, 1:3] <- A11
+   Y <- t(A %*% X)
R> }
\end{CodeInput}
\end{CodeChunk}
We can implement Method \ref{cointegrationRank1} by using the \code{Coint()} function with the argument \code{type = "acf"}, where \code{c0} and \code{m} are set to their default values.
\begin{CodeChunk}
\begin{CodeInput}
R> res <- Coint(Y, type = "acf")
R> print(res)
\end{CodeInput}
\begin{CodeOutput}	
	Cointegration analysis for vector time series

Using acf method
The estimated number of cointegration rank = 3, Time lag = 5
\end{CodeOutput}
\end{CodeChunk}
The estimates $\hat{r}$ and $\hat{\bA}$ can be extracted by the following codes:
\begin{CodeChunk}
\begin{CodeInput}
R> coint_r <- res$coint_rank
R> A_hat <- res$A
\end{CodeInput}
\end{CodeChunk}

 \noindent Method \ref{cointegrationRank2} can be implemented by using the \code{Coint()} function with the argument \code{type = "urtest"}. Additionally, by setting the argument \code{type = "both"} when calling \code{Coint()}, Methods \ref{cointegrationRank1} and \ref{cointegrationRank2} will be applied simultaneously. The results indicate that both methods yield the estimated cointegration rank $\hat{r}=3$.
\begin{CodeChunk}
\begin{CodeInput}
R> Coint(Y, type = "both")
\end{CodeInput}
\begin{CodeOutput}
	Cointegration analysis for vector time series

Using both two methods
The estimated number of cointegration rank:
      acf urtest
r_hat   3      3
Time lag = 5
\end{CodeOutput}
\end{CodeChunk}

\subsection[]{High-dimensional hypothesis testings in \proglang{R}}\label{sec:tstest_fun}
The white noise test and the martingale difference test for high-dimensional time series described in Sections \ref{sec:white noise} and \ref{sec:martingale} are implemented in the \pkg{HDTSA} package through the functions \code{WN_test()} and \code{MartG_test()}, respectively.

The function \code{WN_test()} is used to test whether $H_0$ given in \eqref{eq:wntest} holds, which indicates that $\by_t$ is white noise. The function with its default arguments is:
\begin{Code}
WN_test(Y, lag.k = 2, B = 1000, kernel.type = c("QS", "Par", "Bart"),
       pre = FALSE, alpha = 0.05, control.PCA = list())
\end{Code}
where
\begin{itemize}
	\item{
		\code{Y} is the $n\times p$ data matrix ${\bf Y} = ({\bf y}_1, \dots , {\bf y}_n )^{\T}$, where $\{\by_t\}_{t=1}^n$ are the observations of $p$-dimensional times series.
	}
        \item{
        \code{lag.k} represents the time lag $K$ involved in \eqref{T_n1} with the default value being 2.
        }
        \item{
	    \code{B} represents the number of bootstrap replications $B$ used for generating multivariate normally distributed random vectors when determining the critical value. The default value is 1000.
        }
	\item{
        \code{kernel.type} specifies the kernel function used for calculating the matrix $\bTheta$ when determining the critical value. It can take three values: \code{"QS"} (the default) for the Quadratic spectral kernel, \code{"Par"} for the Parzen kernel, and \code{"Bart"} for the Bartlett kernel.
	}
        \item{
        \code{pre} is a boolean indicating if the time series PCA introduced in Section \ref{sec:pca} should be performed on $\{\by_t\}_{t=1}^{n}$ before implementing the white noise test (see Section \ref{sec:white noise} for more details). The default value is \code{FALSE}. If \code{pre = TRUE}, the PCA will be implemented by using the function \code{PCA_TS()} with the arguments passed by \code{control.PCA}.
        }
        \begin{itemize}
            \item{
            \code{control.PCA} is a list of arguments passed to the function \code{PCA_TS()}, including \code{lag.k}, \code{opt}, \code{thresh}, \code{delta}, and the associated arguments passed to the function \code{clime()} (when \code{opt = 2}).
            }

        \end{itemize}

	\item{
		\code{alpha} represents the significance level $\alpha \in (0,1)$ of the white noise test. The default value is $0.05$.
        }

\end{itemize}

The function \code{MartG_test()} is used to test whether $H_0$ given in \eqref{eq:MDStest} holds, which indicates that $\by_t$ is a martingale difference sequence. 
The function with its default arguments is:
\begin{Code}
MartG_test(Y, lag.k = 2, B = 1000, type = c("Linear", "Quad"), 
          alpha = 0.05, kernel.type = c("QS", "Par", "Bart"))
\end{Code}
The arguments \code{Y}, \code{B}, \code{alpha} and \code{kernel.type} are same as those in the function \code{WN_test()}. \code{lag.k} represents the time lag $K$ involved in \eqref{T_n2} with the default value 2. The argument \code{type} specifies the map $\boldsymbol{\phi}(\cdot)$ specified in Section \ref{sec:martingale}, which can take the values: \code{"Linear"} for linear identity map $\boldsymbol{\phi}(\bx)=\bx$, and \code{"Quad"} for $\boldsymbol{\phi}(\mathbf{x})=\{ {\mathbf{x}^{\T},(\mathbf{x}^2)^{\T}} \}^{\T}$ including both linear and quadratic terms, where $\mathbf{x}^2=(x_1^2,\ldots,x_p^2)^{\T}$ with $\bx = (x_1,\ldots,x_p)^{\T}$. It can also be specified by \proglang{R} users. See \cite{Chang+Jiang+Shao:2022} for more details.
%

Both the \code{WN_test()} and \code{MartG_test()} functions return objects of the class \code{"hdtstest"}. Each object consists of the following parts:
\begin{itemize}
	\item{
		\code{statistic}, the value of the test statistic.
	}
	\item{
		\code{p.value}, the p-value of the test.
	}
\end{itemize}

The implementations of the functions \code{WN_test()} and \code{MartG_test()} involve computationally intensive bootstrap procedures to determine the critical values. For example, in the white noise test introduced in Section \ref{sec:white noise}, the bootstrap procedure requires performing the matrix multiplication of a $p^2K \times p^2K$ matrix $\bI_K \otimes\hat{\boldsymbol{\Omega}}$ with a $(p^2K)$-dimensional vector $\sum_{t=1}^{\tilde{n}}\eta_{i,t}{\bf f}_t$ in each of the $B$ iterations. To improve computational efficiency, we transform these $B$ iterations into a matrix multiplication, which can be efficiently handled using the \pkg{RcppEigen} package. 

We use the following simulated example to show the usage of the \code{WN_test()} and \code{MartG_test()} functions. 

\medskip

\noindent{\bf Example 6.} Let $\by_1,\ldots, \by_n \stackrel{{\rm i.i.d.}}{\sim} \mathcal{N}({\bf 0}, {\bf I}_p)$ with $(n,p)=(200,10)$. Hence, the sequence $\{\by_t\}_{t=1}^n$ is both white noise and a martingale difference sequence.
%
\begin{CodeChunk}
	\begin{CodeInput}
R> library("HDTSA")
R> n <- 200
R> p <- 10
R> set.seed(0)
R> Y <- matrix(rnorm(n * p), n, p)
\end{CodeInput}
\end{CodeChunk}
We can call the \code{WN_test()} function with all arguments set to their default values to implement the white noise test for \eqref{eq:wntest} introduced in Section \ref{sec:white noise}.
\begin{CodeChunk}
\begin{CodeInput}
R> set.seed(0)
R> wn_res1 <- WN_test(Y)
R> print(wn_res1)
	\end{CodeInput}
\begin{CodeOutput}
	Testing for white noise hypothesis in high dimension

Statistic = 3.48 , p-value = 0.12
Time lag = 2
Symmetric kernel = QS
\end{CodeOutput}
\end{CodeChunk}
Furthermore, by setting the argument \code{pre = TRUE} and specifying the argument \code{control.PCA}, we can pre-process the time series $\{\by_t\}_{t=1}^n$ using the PCA before applying the white noise test, which helps to improve its performance.
\begin{CodeChunk}
\begin{CodeInput}
R> set.seed(0)
R> wn_res2 <- WN_test(Y, pre = TRUE, control.PCA = list(thresh = TRUE))
R> print(wn_res2)
\end{CodeInput}
\begin{CodeOutput}
	Testing for white noise hypothesis in high dimension

Statistic = 3.28 , p-value = 0.21
Time lag = 2
Symmetric kernel = QS
\end{CodeOutput}
\end{CodeChunk}
We can call the \code{MartG_test()} function to implement the martingale difference test for \eqref{eq:MDStest} introduced in Section \ref{sec:martingale}. We set the argument \code{type = "Quad"}, which corresponds to the map $\boldsymbol{\phi}(\mathbf{x})=\{ {\mathbf{x}^{\T},(\mathbf{x}^2)^{\T}} \}^{\T}$. 
\begin{CodeChunk}
	\begin{CodeInput}
R> set.seed(0)
R> mds_res1 <- MartG_test(Y, type = "Quad")
R> print(mds_res1)
 \end{CodeInput}
\begin{CodeOutput}
	Testing for martingale difference hypothesis in high dimension

Statistic = 35.5 , p-value = 0.84
Time lag = 2
Symmetric kernel = QS
Data map : Quad
\end{CodeOutput}
\end{CodeChunk}





%
\noindent The test statistic and p-value of the test can be extracted by the following codes:
\begin{CodeChunk}
\begin{CodeInput}
R> statistic <- mds_res1$statistic
R> pv <- mds_res1$p.value
\end{CodeInput}
\end{CodeChunk}
The map $\boldsymbol{\phi}(\cdot)$ can also be provided by the \proglang{R} user. For example, we can use $\boldsymbol{\phi}(\bx)=\cos(\bx)$ to capture certain type of nonlinear dependence, where $\cos(\bx) = (\cos x_1,\ldots,\cos x_p)^{\T}$ for $\bx = (x_1,\ldots,x_p)^{\T}$. The martingale difference sequence test based on this map can be implemented using the following codes:
 \begin{CodeChunk}
 	\begin{CodeInput}
R> map <- cos(Y)
R> set.seed(0)
R> print(mds_res2 <- MartG_test(Y, type = map))
	\end{CodeInput}
\begin{CodeOutput}
	Testing for martingale difference hypothesis in high dimension

Statistic = 5.37 , p-value = 0.84
Time lag = 2
Symmetric kernel = QS
Data map : User define
\end{CodeOutput}
\end{CodeChunk}
The \code{map} defined above is an $n \times p$ matrix, which requires a large amount of memory when $n$ or $p$ is very large. To reduce the memory usage, we can define \code{map} as an \proglang{R} expression instead. This can be accomplished using the built-in \proglang{R} functions \code{quote()}, \code{substitute()}, \code{expression()} and \code{parse()}.
\begin{CodeChunk}
	\begin{CodeInput}
R> map <- quote(cos(Y))
R> map <- substitute(cos(Y))
R> map <- expression(cos(Y))
R> map <- parse(text="cos(Y)")
\end{CodeInput}
\end{CodeChunk}
%

\section{Real data applications} \label{sec:realdata}
In this section, we illustrate the application of the methods in the \pkg{HDTSA} package through two real data sets. 


\subsection{Fama–French return data}

We consider the Fama-French $10 \times 10$ return series available on \url{http://mba.tuck.dartmouth.edu/pages/faculty/ken.french/data_library.html}. The portfolios are constructed by the intersections of 10 levels of size (market equity), denoted by ${\rm S}_1, \ldots, {\rm S}_{10}$, and 10 levels of the book equity to market equity ratio (BE), denoted by ${\rm BE}_1,\ldots,{\rm BE}_{10}$. We collect the monthly returns from January 1964 to December 2021, which contains 69600 observations for total 696 months. The raw data, with the missing values imputed by zeros, is included in the package under the name \code{FamaFrench}. In this dataset, each row represents a monthly observation, and the columns correspond to return series for different sizes and BE-ratios. We have also collected the monthly market return series for the same time period from the same website mentioned above, which is presented in the column named \code{MKT.RF} in the \code{FamaFrench} data.



\begin{CodeChunk}
    \begin{CodeInput}
R> library("HDTSA")
R> data(FamaFrench, package = "HDTSA")
R> head(FamaFrench, c(10L, 9L))
    \end{CodeInput}
    \begin{CodeOutput}
     DATE MKT.RF  S1.BE1  S2.BE1  S3.BE1  S4.BE1  S5.BE1  S6.BE1  S7.BE1
1  196401   2.24 10.6958  4.0064  3.9031  3.5609  4.1416  4.6475  5.3752
2  196402   1.54  1.8582  2.8364 -1.8033  6.9066  4.1488  1.4281  2.9116
3  196403   1.41  3.1559 -2.5998 -1.8414  4.9522  3.9522  1.3513  4.5041
4  196404   0.10 -0.6093  1.1828 -0.2316  0.3536 -2.4791 -0.1026 -1.3847
5  196405   1.42 -2.7943 -3.2814 -1.2380 -0.6127  3.0451  2.4931  0.9711
6  196406   1.27  5.0302  3.0108  0.6005  2.3868 -0.7560  4.0298  0.3568
7  196407   1.74  1.4392  1.2916  3.7764  3.1806  2.8539  5.3740  1.7097
8  196408  -1.44  2.6804  4.0781 -2.8299 -3.6800 -1.7464  1.6888 -1.1816
9  196409   2.69  3.1147  8.6344 11.0110  4.6772  5.5467 10.7354  3.9367
10 196410   0.59  2.5993  4.5398 -0.7264  1.0328  2.3163  4.4641  1.8770
    \end{CodeOutput}
\end{CodeChunk}

\begin{figure}[htbp]
\centering
\includegraphics[width=13cm]{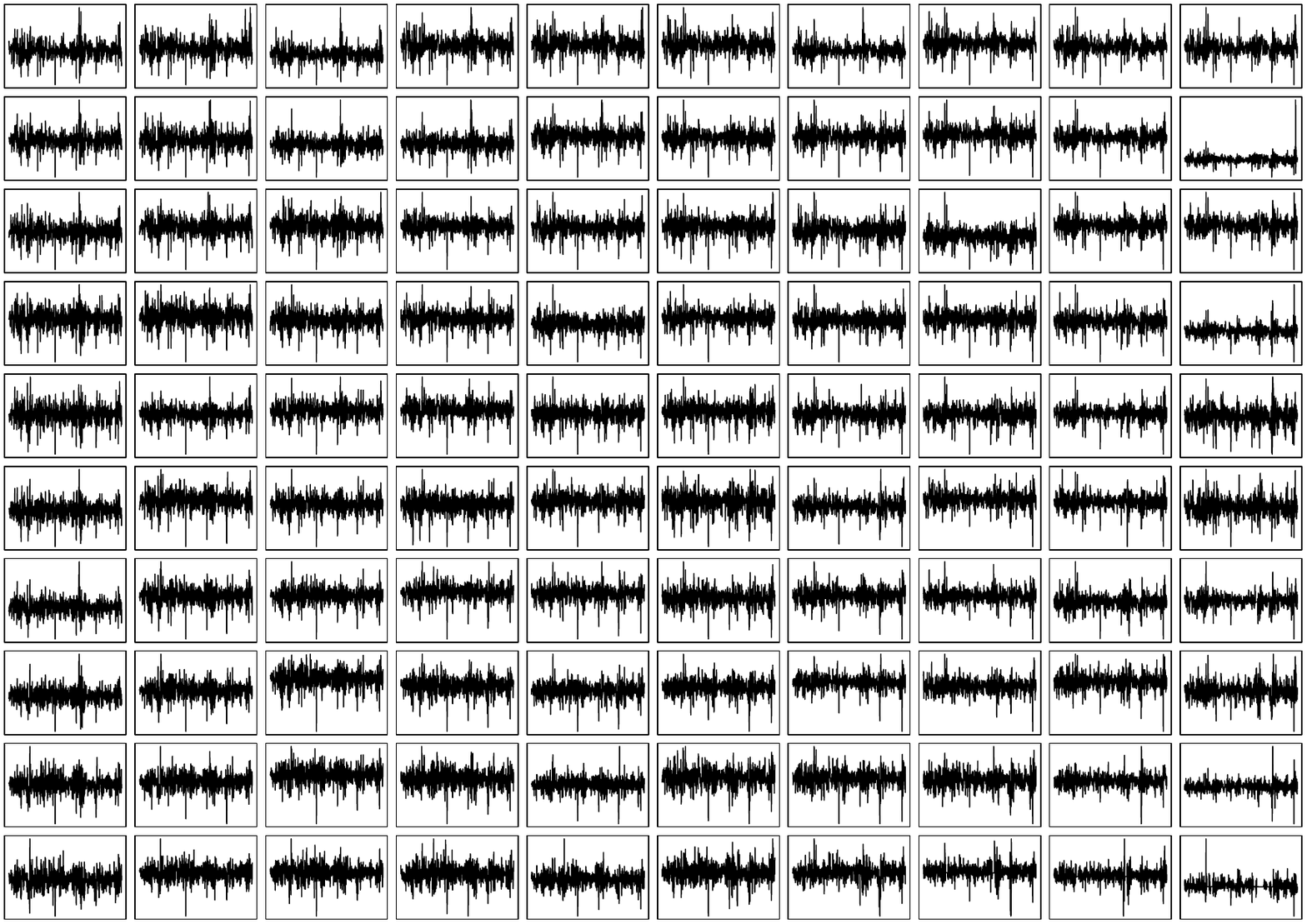}
\caption{\label{fig:FamaFrench} The time series plots of 100 market-adjusted returns formed on different levels of size (by rows) and book equity to market equity ratio (by columns). The horizontal axis represents time and the vertical axis represents the monthly returns.}
\end{figure}

Since all the 100 series are correlated with the overall market condition, following \cite{Wang+Liu+Chen:2019}, we first remove the market effects by fitting a standard CAPM model to each of the series. The data after removing the market effects is named \code{Y_2d}, which represents a $n \times p$ matrix $(\by_1, \ldots, \by_n)^{\T}$ with $\by_t \in \mathbb{R}^p$, $n=696$ and $p=100$. Figure~\ref{fig:FamaFrench} shows the time series plots of the market-adjusted return series. 
\begin{CodeChunk}
    \begin{CodeInput}
R> reg <- lm(as.matrix(FamaFrench[, -c(1:2)]) ~ as.matrix(FamaFrench$MKT.RF))
R> Y_2d = reg$residuals
    \end{CodeInput}
\end{CodeChunk}

\noindent \textit{Factor Modeling}

The 100 return series can be represented as a 100-dimensional vector time series $\by_t$, which can be analyzed using the factor model introduced in Section \ref{sec:factors}. We first run the \code{Factors()} function with the argument \code{lag.k = 5} to fit the factor model following the standard estimation procedures in Section \ref{sec:factors}. The estimated number of factors is $\hat{r}=1$.

\begin{CodeChunk}
    \begin{CodeInput}
R> res_factors <- Factors(Y_2d, lag.k = 5)
R> print(res_factors)
    \end{CodeInput}
\begin{CodeOutput}
	Factor analysis for vector time series

The estimated number of factors = 1
Time lag = 5
\end{CodeOutput}
\end{CodeChunk}
Then, we can use the following codes to retrieve the estimated $100 \times 1$ loading matrix $\hat{\bA}$, which is named \code{est_loading_Fac}.
\begin{CodeChunk}
    \begin{CodeInput}  
R> est_loading_Fac <- res_factors$loading.mat
R> dim(est_loading_Fac)
\end{CodeInput}
\begin{CodeOutput}
[1] 100   1
\end{CodeOutput}
\begin{CodeInput}
R> head(round(est_loading_Fac, 5))
    \end{CodeInput}
\begin{CodeOutput}
         [,1]
[1,] -0.14325
[2,] -0.13461
[3,] -0.11922
[4,] -0.15048
[5,] -0.17276
[6,] -0.13638
\end{CodeOutput}
\end{CodeChunk}
As mentioned in Section \ref{sec:factors}, since $\hat{\bA}$ is an orthogonal matrix, the univariate latent factor process $x_t$ can be estimated by $\hat{x}_t = \hat{\bA}^{\T}\by_t$. We can get $\hat{x}_t$ by
\begin{CodeChunk}
\begin{CodeInput}
R> est_ts_Fac <- res_factors$X
\end{CodeInput}
\end{CodeChunk}
Hence, in order to predict $\by_t$, we only need to fit a univariate time series model for $\hat{x}_t$ by calling the function \code{auto.arima()} in the \pkg{forecast} package. The one-step ahead prediction $\hat{x}_{n+1}$ can be obtained using the function \code{predict()}. Then we have $\hat{\by}_{n+1} = \hat{\bA} \hat{x}_{n+1}$. 
\begin{CodeChunk}
\begin{CodeInput}
R> fit_Fac <- forecast::auto.arima(est_ts_Fac, ic = "aic")
R> pred_Fac <- predict(fit_Fac, n.ahead = 1)
R> pred_Fac$pred    
\end{CodeInput}
\begin{CodeOutput}
Time Series:
Start = 697 
End = 697 
Frequency = 1 
[1] 6.21131
\end{CodeOutput}
\begin{CodeInput}
R> pred_fac_Y1 <- as.matrix(pred_Fac$pred) %*% t(est_loading_Fac)
R> head(round(pred_fac_Y1, 5), c(1L, 8L))
\end{CodeInput}
\begin{CodeOutput}
         [,1]     [,2]     [,3]     [,4]    [,5]     [,6]     [,7]    [,8]
[1,] -0.88974 -0.83608 -0.74051 -0.93468 -1.0731 -0.84712 -1.14007 -1.0598
\end{CodeOutput}
\end{CodeChunk}

For ease of use, we have also included \code{predict}
method for the result fitted by \code{Factors()} in the \pkg{HDTSA} package, to make predictions based on the above process. Hence, we can also get $\hat{\by}_{n+1}$ by running the following codes directly. See the manual and the help file of \code{predict.factors()} for usage details and examples.

\begin{CodeChunk}
\begin{CodeInput}
R> pred_fac_Y2 <- predict(res_factors, n.ahead = 1)
R> head(pred_fac_Y2, c(1L, 8L))
\end{CodeInput}
\begin{CodeOutput}
           [,1]     [,2]     [,3]     [,4]    [,5]     [,6]     [,7]    [,8]
1 step -0.88974 -0.83608 -0.74051 -0.93468 -1.0731 -0.84712 -1.14007 -1.0598
\end{CodeOutput}
\end{CodeChunk}

\medskip

\noindent \textit{Principal Component Analysis}

We can also apply the time series PCA introduced in Section \ref{sec:pca} to the 100-dimensional vector return time series $\by_t$. The latent segmentation structure of $\by_t$ can be estimated by calling the \code{PCA_TS()} function, leading to 96 univariate time series and one 4-dimensional time series.
\begin{CodeChunk}
    \begin{CodeInput}
R> res_pca <- PCA_TS(Y_2d, lag.k = 5, thresh = TRUE)
R> print(res_pca)
    \end{CodeInput}
\begin{CodeOutput}
	Principal component analysis for time series

Maximum cross correlation method
The number of groups = 97
The numbers of members in groups containing at least two members: 4
\end{CodeOutput}
\end{CodeChunk}
The transformed series $\hat{\bx}_t:=(\hat{x}_{t,1},\ldots,\hat{x}_{t,100})^{\T}$ and the segmentation structure of $\hat{\bx}_t$ can be obtained by the following codes. The 100 component series of $\hat{\bx}_t$ are segmented into 97 uncorrelated groups: $\hat{x}_{t,1}$, $\hat{x}_{t,2}$, $\hat{x}_{t,63}$ and $\hat{x}_{t,96}$ form Group 1, while each of the remaining component series of $\hat{\bx}_t$ forms its own individual group. 
\begin{CodeChunk}
    \begin{CodeInput}
R> est_X <- res_pca$X
R> X_group <- res_pca$Groups
R> head(X_group, c(6L, 9L))
    \end{CodeInput}
\begin{CodeOutput}
     Group 1 Group 2 Group 3 Group 4 Group 5 Group 6 Group 7 Group 8 Group 9
[1,]       1       3       4       5       6       7       8       9      10
[2,]       2       0       0       0       0       0       0       0       0
[3,]      63       0       0       0       0       0       0       0       0
[4,]      96       0       0       0       0       0       0       0       0
\end{CodeOutput}
\end{CodeChunk}

\begin{figure}[htbp]
\centering
\includegraphics[width = 13cm]{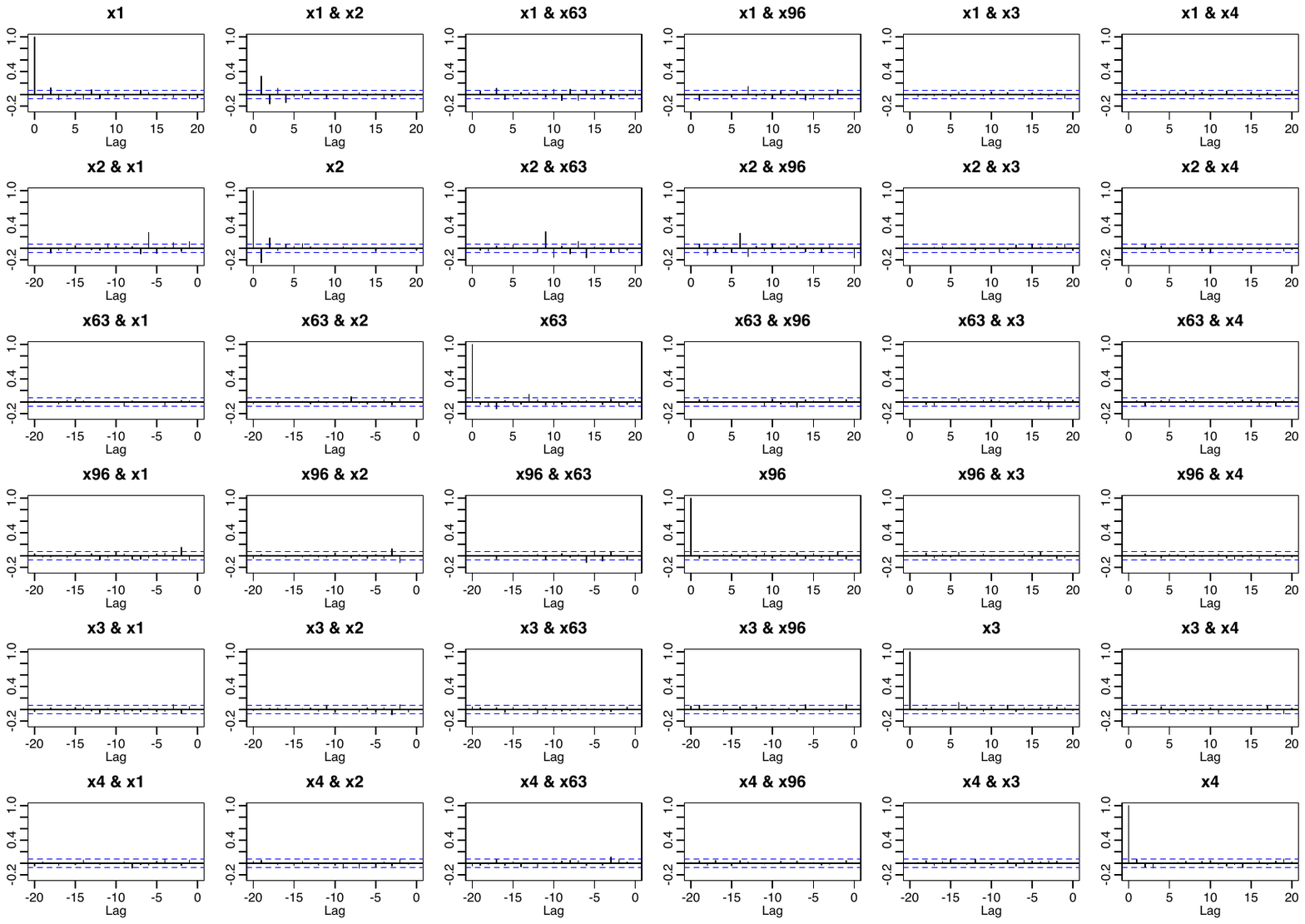}
\caption{\label{fig:pca_realdata} Cross correlogram of the 6 component series of $\hat{\bx}_t$ in the Groups 1--3.}
\end{figure}

Figure~\ref{fig:pca_realdata} plots the cross correlations of the 6 component series of $\hat{\bx}_t$ in Groups 1--3. It shows that only the 4 component series in Group 1 are significantly correlated at some time lag. There is little significant cross correlation among all the other pairs of component series.

Since there are no cross correlations among different groups at all time lags, those 97 groups can be analyzed separately. We then forecast the components of $\hat{\bx}_t$ according to the segmentation, that is, one 4-dimensional VAR model for the first group and a univariate AR model for each of the other 96 groups. 
\begin{CodeChunk}
    \begin{CodeInput}
R> X_pred <- matrix(0, 1, 100)
R> group_1 <- est_X[, X_group[, "Group 1"]]
R> colnames(group_1) <- c("x1", "x2", "x63", "x96")
R> var_group_1 = vars::VAR(group_1, type = "const", lag.max = 6, ic = "AIC")
R> pre_obj = predict(var_group_1, n.ahead = 1)$fcst
R> X_pred[ , X_group[, "Group 1"]] <- cbind(pre_obj$x1[1], pre_obj$x2[1],
+                                           pre_obj$x63[1], pre_obj$x96[1])
R> for (jj in 2:97){
+    group_jj <- est_X[, X_group[, jj]]
+    group_jj_fit  <- forecast::auto.arima(group_jj, max.q = 0, max.d = 0,
+                                          ic = "aic")
+    pre_obj <- predict(group_jj_fit, n.ahead = 1)$pred
+    X_pred[ , X_group[, jj]] <- pre_obj
+  }
    \end{CodeInput}
\end{CodeChunk}
In the codes shown above, \code{X_pred} represents the one-step ahead forecast $\hat{\bx}_{n+1}$. The forecasted value for $\by_{n+1}$ is obtained via the transformation $\hat{\by}_{n+1} = \hat{\bB}^{-1}\hat{\bx}_{n+1}$, where $\hat{\bf B}=\hat{\bf \Gamma}^{\T}\hat{\bf V}^{-1/2}$ with $\hat{\bf \Gamma}$ specified in Section \ref{sec:pca} and $\hat{\bV}^{-1}$ being the estimated precision matrix of $\by_t$.
\begin{CodeChunk}
    \begin{CodeInput}
R> pred_pca_Y1 <- X_pred %*% t(solve(res_pca$B))
R> head(round(pred_pca_Y1, 5), c(1L, 8L))
    \end{CodeInput}
\begin{CodeOutput}
         [,1]     [,2]     [,3]     [,4]    [,5]     [,6]    [,7]     [,8]
[1,] -1.72283 -1.23158 -1.26604 -1.43991 -0.9564 -0.91185 -0.8238 -0.57155
\end{CodeOutput}
\end{CodeChunk}

The above forecasting procedure can be implemented by the \code{predict} method in the \pkg{HDTSA} package directly. See the manual and the help file of \code{predict.tspca()} for more details and examples. 


\begin{CodeChunk}
    \begin{CodeInput}
R> pred_pca_Y2 <- predict(res_pca, n.ahead = 1)
    \end{CodeInput}
\end{CodeChunk}

\medskip

\noindent \textit{Matrix Factor Model}

The data \code{Y_2d} can also be rearranged into a 3-dimensional data array \code{Y}, where \code{Y[t, , ]} is the $10\times 10$ matrix $\bY_t$ for each $t$, with the $i$th row representing size level ${\rm S}_i$ and the $j$th column representing BE-ratio level ${\rm BE}_j$. 
\begin{CodeChunk}
    \begin{CodeInput}
R> Y = array(NA, dim = c(NROW(Y_2d), 10, 10))
R> for (tt in 1:NROW(Y_2d)) {
+    for (ii in 1:10) {
+      Y[tt, ii, ] <- Y_2d[tt, (1 + 10*(ii - 1)):(10 * ii)]
+    }
+  }
    \end{CodeInput}
\end{CodeChunk}

Then we can analyze the matrix time series $\bY_t$ using the CP-decomposition introduced in Section \ref{sec:CP}. Using the refined estimation method (Method \ref{mtd:refined}),
we get $\hat{d}=1$. 
\begin{CodeChunk}
    \begin{CodeInput}
R> res_cp <- CP_MTS(Y, lag.k = 20, method = "CP.Refined")
R> print(res_cp)
    \end{CodeInput}
\begin{CodeOutput}
	Estimation of matrix CP-factor model

Method: CP.Refined
The estimated number of factors d = 1
\end{CodeOutput}
\end{CodeChunk}
The loading matrices $\hat{\bA}$, $\hat{\bB}$ and the univariate latent time series $\hat{x}_t$ can be obtained as follows. 
\begin{CodeChunk}
    \begin{CodeInput}  
R> est_ts_cp <- res_cp$f
R> est_loading_cp_A <- res_factors$A
R> est_loading_cp_B <- res_factors$B
R> A_B <- cbind(est_loading_cp_A, est_loading_cp_B)
R> colnames(A_B) <- c("A", "B")
R> print(round(A_B, 5))
    \end{CodeInput}
\begin{CodeOutput}
             A        B
 [1,] -0.55402 -0.29145
 [2,] -0.42856 -0.32377
 [3,] -0.37482 -0.29770
 [4,] -0.34046 -0.31765
 [5,] -0.28324 -0.30743
 [6,] -0.25900 -0.28718
 [7,] -0.21067 -0.29558
 [8,] -0.18795 -0.31388
 [9,] -0.15203 -0.34249
[10,] -0.05359 -0.37486
\end{CodeOutput}
\end{CodeChunk}
%

Hence, in order to predict $\bY_{n+1}$, we fit a univariate time series model to $\{\hat{x}_t\}_{t=1}^n$ by calling the \code{auto.arima} function in the \pkg{forecast} package, and obtain the forecast $\hat{x}_{n+1}$. Then, we have $\hat{\bY}_{n+1} = \hat{x}_{n+1}\hat{\bA} \hat{\bB}^{\T}$.
\begin{CodeChunk}
    \begin{CodeInput}  
R> arima_cp <- forecast::auto.arima(est_ts_cp, ic = "aic")
R> pred_cp <- predict(arima_cp, n.ahead = 1)$pred
R> pred_cp
    \end{CodeInput}
\begin{CodeOutput}
Time Series:
Start = 697 
End = 697 
Frequency = 1 
[1] -6.811228
\end{CodeOutput}
    \begin{CodeInput}
R> pred_cp_Y1 <- est_loading_cp_A %*% diag(pred_cp,1,1) %*% t(est_loading_cp_B)
    \end{CodeInput}
\end{CodeChunk}

The above forecasting procedure can be implemented by the  \code{predict} method in the \pkg{HDTSA} package directly. See the manual and the help file of \code{predict.mtscp()} for more details and examples.


\begin{CodeChunk}
    \begin{CodeInput}
R> pred_cp_Y2 <- predict(res_cp, n.ahead = 1)[[1]]
    \end{CodeInput}
\end{CodeChunk}

\subsection{Industrial Production indices data}

We consider 7 U.S. Industrial Production indices from January 1947 to December 2023, published by the U.S. Federal Reserve. The data is included in the \pkg{HDTSA} package under the name \code{IPindices}, with columns named \code{INDPRO}, \code{IPB54000S}, \code{IPFIN}, \code{IPMANSICS}, \code{IPMAT}, \code{IPMINE}, and \code{IPUTIL}, representing 7 U.S. Industrial Production indices: \textit{the total index}, \textit{nonindustrial supplies}, \textit{final products}, \textit{manufacturing}, \textit{materials}, \textit{mining}, and \textit{utilities}, respectively.
\begin{CodeChunk}
    \begin{CodeInput}
R> library("HDTSA")
R> data(IPindices, package = "HDTSA")
R> head(IPindices)
    \end{CodeInput}
    \begin{CodeOutput}
        DATE  INDPRO IPB54000S IPFINAL IPMANSICS   IPMAT  IPMINE IPUTIL
1 1947-01-01 13.6554   16.2061 13.9138   13.1480 12.3704 47.3987 6.2503
2 1947-02-01 13.7361   16.4927 13.9413   13.1990 12.4680 47.9482 6.3011
3 1947-03-01 13.8167   16.5563 13.9413   13.2501 12.8828 48.8412 6.3519
4 1947-04-01 13.7092   16.6519 13.9138   13.2756 12.5656 44.7196 6.4790
5 1947-05-01 13.7629   16.7792 13.8862   13.1735 12.6144 48.4291 6.5806
6 1947-06-01 13.7629   16.5882 13.9138   13.1735 12.4924 48.0170 6.6060
    \end{CodeOutput}
\end{CodeChunk}

\begin{figure}[htbp]
\centering
\includegraphics[width=13cm]{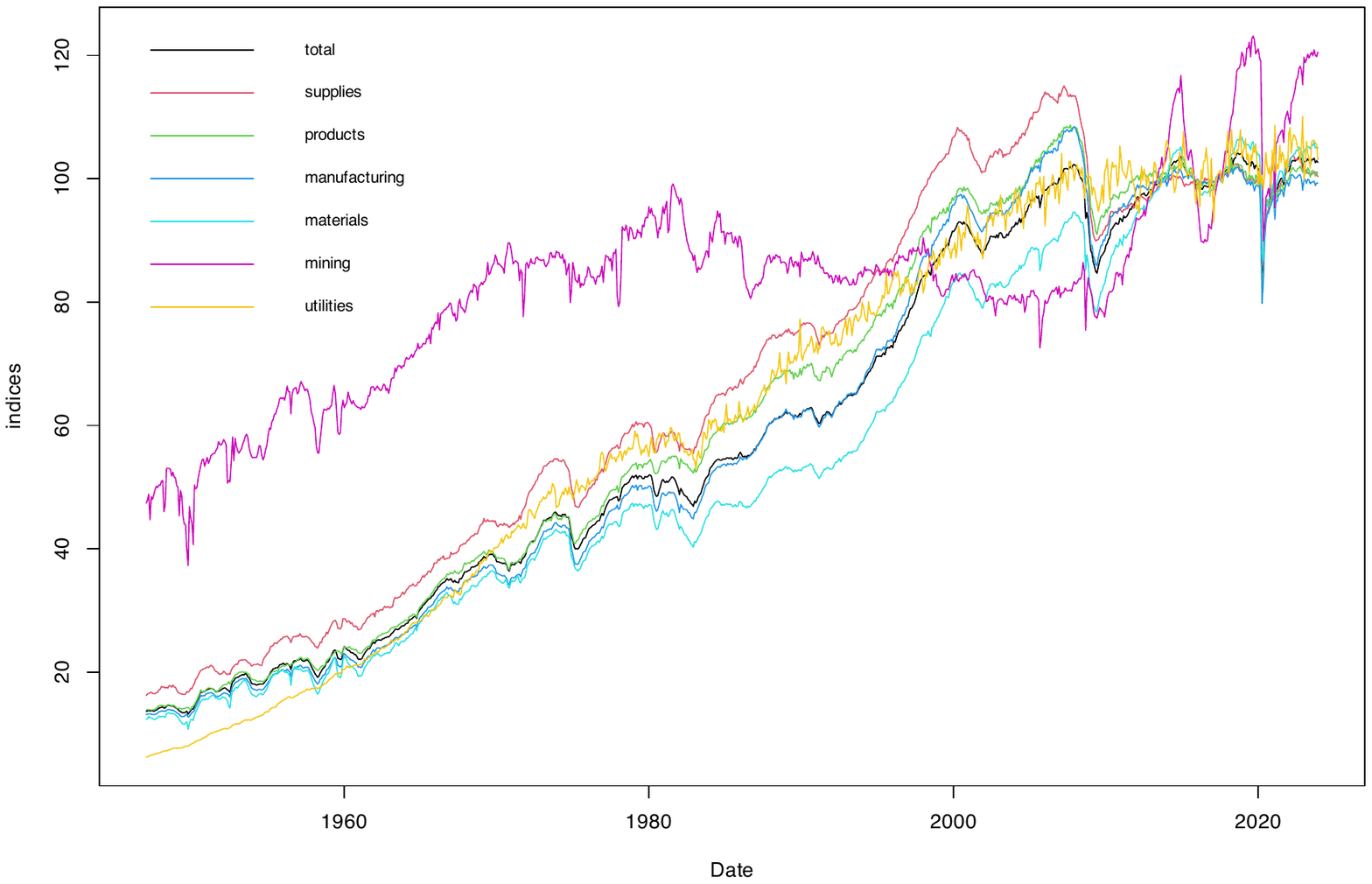}
\caption{\label{fig:IPindices} Time series plots of the 7 monthly U.S. Industrial Production indices in January 1947-December 2023.}
\end{figure}

The original 7 Industrial Production index time series are plotted in Figure~\ref{fig:IPindices}. It can be observed that all 7 time series are non-stationary. Therefore, we consider investigating whether a cointegration relationship exists among these 7 time series. Denote by $\by_t$ a 7-dimensional vector time series consisting of these 7 time series. If cointegration is identified, we can then fit a vector error correction model (VECM) for $\by_t$.
%

\begin{figure}[htbp]
\centering
\includegraphics[width=11.5cm]{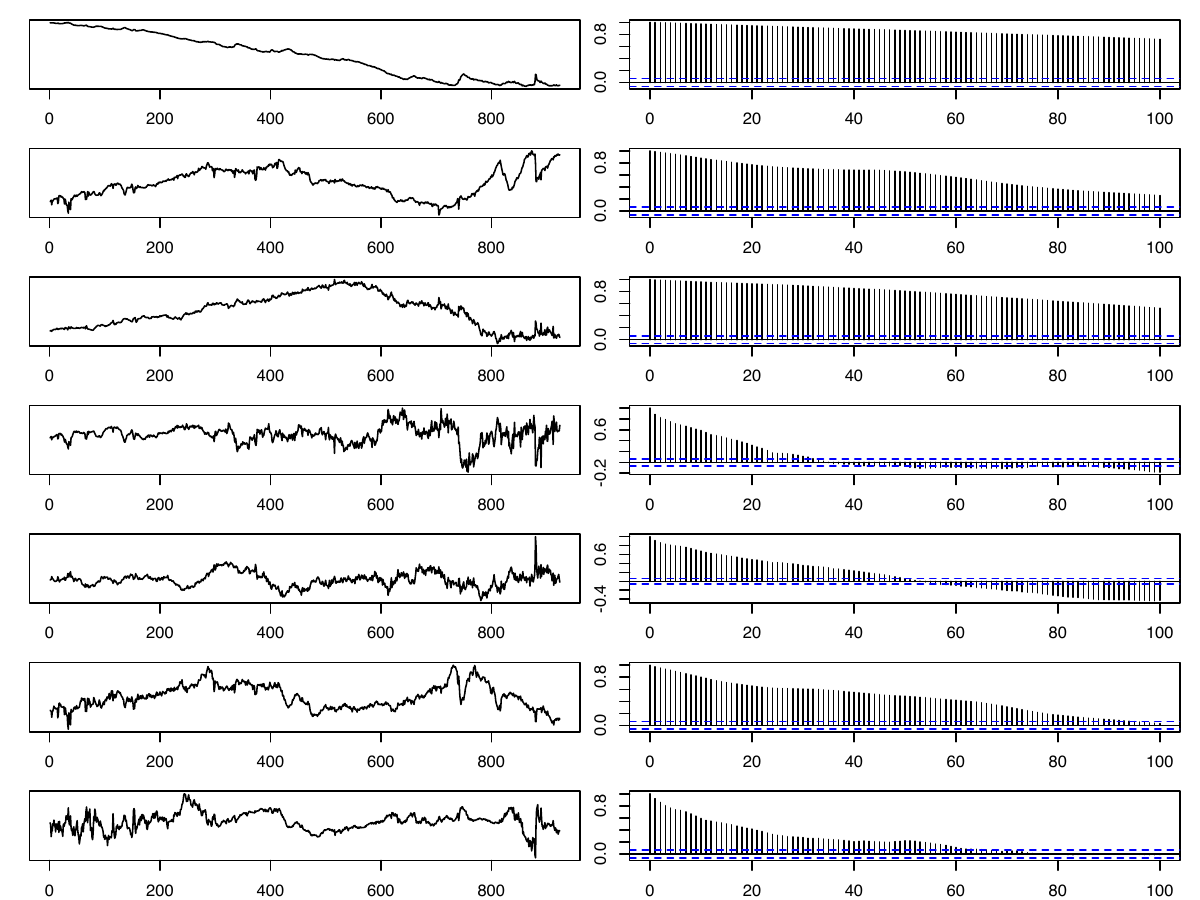}
\caption{\label{fig:acf_indices} Time series plots of the estimated $\hat{\bx}_t$ and their sample ACF for the 7 monthly U.S. Industrial Production Indices.}
\end{figure}

We first determine the cointegration rank $r$ using Method \ref{cointegrationRank2} 
by calling the \code{Coint()} function with arguments \code{lag.k = 10} and \code{type = "urtest"}. The estimated cointegration rank is $\hat{r}=4$.
\begin{CodeChunk}
    \begin{CodeInput}
R> data <- IPindices[, -1]
R> res_coint <- Coint(data, lag.k = 10, type = "urtest")
R> print(res_coint)
    \end{CodeInput}
\begin{CodeOutput}
	Cointegration analysis for vector time series

Using unit root test
The estimated number of cointegration rank = 4, Time lag = 10
\end{CodeOutput}
\end{CodeChunk}
 The estimation of $\bA$ in \eqref{eq:cointegration}, denoted by $\hat{\bA}$, can be obtained by running the following command. The last 4 columns of $\hat{\bA}$ are the loadings of the 4 identified cointegrated variables. 
\begin{CodeChunk}
    \begin{CodeInput}
R> coint_A <- res_coint$A
    \end{CodeInput}
\end{CodeChunk}
Then, we can calculate $\hat{\bx}_{t} = \hat{\bA}^{\T} \by_t$.
\begin{CodeChunk}
\begin{CodeInput}
R> xt <- as.matrix(data) %*% coint_A
R> head(round(xt, 5))
\end{CodeInput}
\begin{CodeOutput}
          [,1]     [,2]    [,3]     [,4]     [,5]    [,6]     [,7]
[1,] -37.84261 37.57578 3.80670 19.10454 -5.33612 1.53287  0.02560
[2,] -38.16756 38.02830 3.94285 19.36786 -5.25153 1.51478  0.00726
[3,] -38.56862 38.87927 3.77272 19.60010 -5.18877 1.49827 -0.10741
[4,] -37.83968 35.01101 3.64518 18.43460 -4.73148 1.28601 -0.05429
[5,] -38.51098 38.48188 4.12887 19.40927 -5.02234 1.47895 -0.05194
[6,] -38.34407 38.12424 4.08260 19.18719 -5.12144 1.51051  0.01555
\end{CodeOutput}
\end{CodeChunk}
The transformed series $\hat{\bx}_{t}$ are plotted in Figure~\ref{fig:acf_indices} together with their sample ACF. As shown in Figure~\ref{fig:acf_indices}, the last 4 series of $\hat{\bx}_{t}$ are all $I(0)$ series, indicating that they are cointegrating errors of $\by_t$. See more discussions in \cite{Zhang+Robinson+Yao:2019}. 

Then, we fit a VECM with the cointegration rank $\hat{r}=4$ by calling the \code{VECM()} function in the \pkg{tsDyn} package.
\begin{CodeChunk}
    \begin{CodeInput}
R> install.packages("tsDyn")
R> library("tsDyn")
R> vecm_model <- VECM(data, lag = 2, r = res_coint$coint_rank,
+                    include = "const", estim = "ML")
    \end{CodeInput}
\end{CodeChunk}
We can obtain the residuals of the model, and then apply the white noise test introduced in Section \ref{sec:white noise} on them. The results show that the 7-dimensional residual vector is white noise, which indicates that the model is well-specified.
\begin{CodeChunk}
    \begin{CodeInput}
R> residuals <- vecm_model$residuals
R> set.seed(0)
R> res_wn <- WN_test(residuals)
R> print(res_wn)
    \end{CodeInput}
\begin{CodeOutput}
	Testing for white noise hypothesis in high dimension

Statistic = 1.55 , p-value = 1
Time lag = 2
Symmetric kernel = QS
\end{CodeOutput}
\end{CodeChunk}

In addition, we can also apply the martingale difference test introduced in Section \ref{sec:martingale} on the residuals. It shows that the residual vector is also a martingale difference sequence. 
\begin{CodeChunk}
    \begin{CodeInput}
R> set.seed(0)
R> res_martg <- MartG_test(residuals)
R> print(res_martg)
    \end{CodeInput}
\begin{CodeOutput}
	Testing for martingale difference hypothesis in high dimension

Statistic = 11.5 , p-value = 0.97
Time lag = 2
Symmetric kernel = QS
Data map : Linear
\end{CodeOutput}
\end{CodeChunk}

\section{Conclusion} \label{sec:Concusion}

In this paper, we present the \proglang{R} package \pkg{HDTSA}, which provides a general framework for analyzing high-dimensional time series data. This package focuses on modeling and statistical inference methods, which are two primary tasks in high-dimensional time series analysis. For high-dimensional time series modeling, the package includes four dimension reduction methods: (a) factor model for vector time series, (b) PCA for vector time series, (c) CP-decomposition for matrix time series, and (d) cointegration analysis for vector time series.  Additional tools for forecasting are also provided, helping \proglang{R} users to make reliable predictions based on these models. For statistical inference, the \pkg{HDTSA} package implements two recently proposed methods for high-dimensional time series: (e) white noise test, and (f) martingale difference test. These methods have rigorous statistical theory guarantees and exhibit good performances in real-world applications. Furthermore, to improve computational efficiency, particularly for high-dimensional time series data, the \pkg{HDTSA} package interfaces with \proglang{C++} through the \pkg{RcppEigen} package. 

The \pkg{HDTSA} package provides powerful and efficient tools for high-dimensional time series analysis. Future extensions of the package may include new methods for analyzing high-dimensional time series, such as statistical inference techniques for high-dimensional spectral density matrices \citep{Chang+Jiang+Tucker+Shao:2022}. Another direction to expand the package would be to further improve the efficiency of the existing methods in the current version. For instance, we plan to develop a new white noise test that not only accelerates computational speed but also reduces memory usage. Incorporating visualization tools, such as plots of factor structures and segmentation patterns, would also benefit users by providing intuitive insights into the data. Finally, providing seamless integration with other time series analysis packages and developing distributed computing algorithms for high-dimensional time series models could make the \pkg{HDTSA} package a more versatile and indispensable tool for \proglang{R} users.

\section*{Computational details}
The results in this paper were obtained using \proglang{R}~4.1.0 on a macOS platform with an M1 Pro CPU. \proglang{R} itself
and all packages used are available from the Comprehensive
\proglang{R} Archive Network (CRAN) at
\url{https://CRAN.R-project.org/}.



\bibliography{refs_arxiv}

\end{document}